\documentclass[journal]{IEEEtran}
\usepackage{cite}

\ifCLASSINFOpdf
  \usepackage[pdftex]{graphicx}
  \graphicspath{{fig/}}
  \DeclareGraphicsExtensions{.pdf,.jpeg,.png}
\else
  \usepackage[dvips]{graphicx}
  \graphicspath{{figs/}}
  \DeclareGraphicsExtensions{.pdf,.jpeg,.png}
\fi

\usepackage{amsmath, amssymb, bm}
\usepackage{textcomp}
\usepackage{gensymb}
\DeclareMathOperator*{\argmax}{argmax}
\DeclareMathOperator*{\argmin}{argmin}
\DeclareMathOperator{\E}{\mathbb{E}}

\begin{document}
% ------------------------------------------------------------------------------
\title{Evaluating Adversarial Evasion Attacks in the Context of Wireless
Communications}

% ------------------------------------------------------------------------------
\author{Bryse Flowers, R. Michael Buehrer, and William C. Headley% 
\thanks{The work of Bryse Flowers was supported in part by the Bradley Masters
Fellowship through the Bradley Department of Electrical and Computer Engineering
at Virginia Tech.}
\thanks{The authors are with the Bradley Department of Electrical and Computer
Engineering, Virginia Tech, Blacksburg, VA, 24061 USA (e-mail: \{brysef,
buehrer, cheadley\}@vt.edu).}
}

% ------------------------------------------------------------------------------
\maketitle

% ------------------------------------------------------------------------------
\begin{abstract}
Recent advancements in radio frequency machine learning (RFML) have demonstrated
the use of raw in-phase and quadrature (IQ) samples for multiple spectrum 
sensing tasks.  Yet, deep learning techniques have been shown, in other
applications, to be vulnerable to adversarial machine learning (ML) techniques, 
which seek to craft small perturbations that are added to the input to cause a
misclassification.  The current work differentiates the threats that adversarial
ML poses to RFML systems based on where the attack is executed from: direct
access to classifier input, synchronously transmitted over the air (OTA), or
asynchronously transmitted from a separate device.  Additionally, the current
work develops a methodology for evaluating adversarial success in the context of
wireless communications, where the primary metric of interest is bit error rate 
and not human perception, as is the case in image recognition.  The methodology
is demonstrated using the well known Fast Gradient Sign Method to evaluate the
vulnerabilities of raw IQ based Automatic Modulation Classification and 
concludes RFML is vulnerable to adversarial examples, even in OTA attacks.
However, RFML domain specific receiver effects, which would be encountered in an
OTA attack, can present significant impairments to adversarial evasion.
\end{abstract}

% ------------------------------------------------------------------------------

% ------------------------------------------------------------------------------
\section{Introduction} \label{sec:intro}
% ------------------------------------------------------------------------------
The advent of deep learning has changed the face of many fields of research
in recent years, including the wireless communications domain.  In particular,
Radio Frequency Machine Learning (RFML), a research thrust championed by DARPA
that seeks to develop RF systems that learn from raw data rather than 
hand-engineered features, has garnered the interest of many researchers.
One subset of RFML deals with utilizing raw in-phase and quadrature
(IQ) samples for spectrum sensing.  Spectrum sensing can be used in Dynamic 
Spectrum Access (DSA) systems to determine the presence of primary and secondary
users in order to adapt transmission parameters to the environment \cite{RN105}
and has obvious applications to signals intelligence.  Prior approaches to 
spectrum sensing were likelihood or feature based
\cite{RN104, RN86, RN87, RN88, RN89, RN90} while more recent approaches leverage
the advances in deep neural networks (DNN) to operate directly on raw IQ samples
\cite{RN27, RN28, RN91, RN92, RN93, Wong:SEI}.

While the popularity of RFML has increased, the study of the vulnerabilities of 
these systems to adversarial machine learning \cite{RN7} has lagged behind. 
Adversarial machine learning consists of learning to apply small perturbations to
input examples that cause a misclassification.  The increased activity in deep
learning research in wireless is sure to draw the attention of attackers in this
domain but is just beginning to be researched \cite{RN100, RN44}.  Adversarial
machine learning could be used, in the context of RFML, to disrupt DSA systems
through primary user emulation \cite{RN2}, evade mobile transmitter tracking
\cite{RN107}, or avoid demodulation by confusing an Automatic Modulation
Classification (AMC) system \cite{RN104}.

While research thrusts towards adversarial machine learning evasion attacks and
defenses can build off of the large body of literature present in the Computer
Vision (CV) domain, RFML has additional adversarial goals and capabilities 
beyond those typically considered in CV.  Adversarial goals must be split
between attacks that have direct access to the classifier input, those that
originate from the transmitter and therefore propagate synchronously with the
underlying transmission through a stochastic channel, and those that originate
asynchronously from a separate transmitter and are only combined at the receiver
or eavesdropper.  Additionally, in the context of wireless communications,
attacks must be characterized against the primary metric of interest, bit error
rate (BER).  An adversary may seek to evade an eavesdropping classifier but that
is of limited benefit if it also corrupts the transmission to a cooperative
receiver.

The current work consolidates the additional adversarial goals and capabilities
present in RFML and proposes a new threat model.  Using the well known Fast
Gradient Sign Method (FGSM) \cite{RN68}, results are presented from multiple
example attacks against raw-IQ based AMC in order to draw general conclusions
about the current vulnerabilities of RFML systems to adversarial machine 
learning attacks that have direct access to the AMC input as well as attacks 
that occur over the air (OTA).  The current work is organized as follows:
Section \ref{sec:related} surveys the related work in this area, Section
\ref{sec:threat} presents a consolidated threat model for RFML systems, Section
\ref{sec:methodology} describes the methodology for executing and evaluating
the adversarial evasion attacks in the context of wireless communications,
Section \ref{sec:performance} and \ref{sec:receiver} analyze the attack's
effectiveness with direct access to the classifier input and in an OTA
environment respectively, and conclusions are presented in Section
\ref{sec:conclusion}.

% Really want this on the second page
\begin{figure*}[ht!]
    \includegraphics[width=\linewidth]{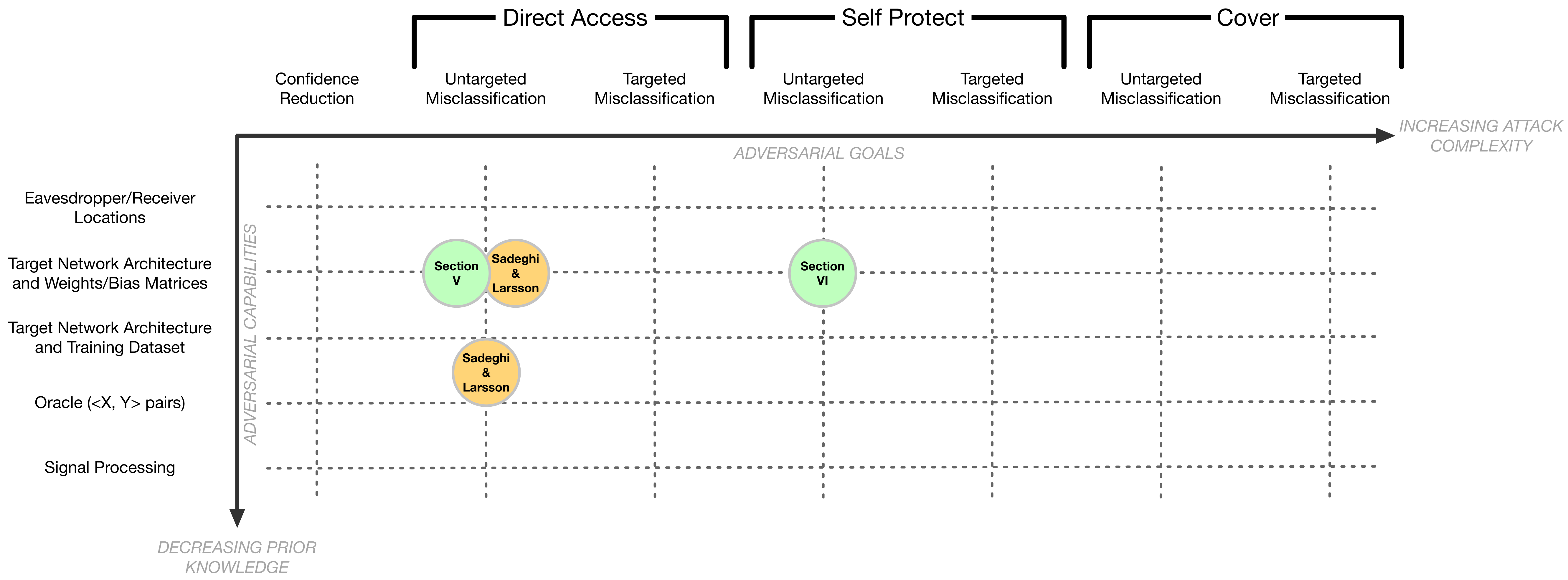}
    \caption{Threat Model for RFML signal classification systems presented in the style of \cite{RN13}.  The current work presents results for untargeted misclassification in both a direct access and self protect scenario with  full knowledge of the target network architecture and learned parameters.   The related work by Sadeghi and Larsson \cite{RN100} presented an analysis of two untargeted misclassification attacks against AMC without a channel model applied to the perturbations.  One attack assumed perfect knowledge of the target network and the other only assumed knowledge of the entire training dataset.}
    \label{fig:threat:ThreatModel}
\end{figure*}

% ------------------------------------------------------------------------------
\section{Related Work} \label{sec:related}
% ------------------------------------------------------------------------------
Threats to machine learning have a wide span in the literature.  Causative
attacks exert influence over the training process to inject vulnerabilities into
the trained classifier \cite{RN78, RN72}.  Exploratory attacks \cite{RN7} seek to
learn information about the classifier.  The current work is primarily concerned
with evasion attacks \cite{RN80, RN25, RN75} which seek to cause a
misclassification at inference time.  Specifically, this work uses the well known
FGSM attack, first proposed in \cite{RN68} for a CV application, as the algorithm
for crafting adversarial perturbations due to its low computational complexity;
however, the methodology for evaluating the attack effectiveness will hold for
all current evasion attacks.

Prior security threats to cognitive signal classifiers have been researched
\cite{RN103, RN96}, yet, the state of the art signal classification systems
use deep learning techniques \cite{RN27, RN28, RN91, RN92, RN93, Wong:SEI}
whose vulnerabilities have not been studied extensively in the context of RF.
In \cite{RN44} and \cite{RN102}, the authors consider adversarial machine
learning for intelligently jamming a deep learning enabled transmitter, at
transmission time and sensing time, to prevent a transmission.  Their work
considers learning OTA by observing an acknowledgement from a receiver as a
binary feedback.  While their work is primarily concerned with
preventing transmission, the current work is primarily concerned with enabling
transmission while avoiding eavesdroppers and is thus fundamentally different.

The work presented in \cite{RN100} is the closest analogy to the current work.  
The authors present a study of a similar neural network architecture \cite{RN27}
using the RML2016.10A dataset \cite{RN29}.  The authors present results from
attacks on this DNN using modifications of FGSM \cite{RN68} and Universal
Adversarial Perturbations (UAP) \cite{RN80}.  Using their adaptation of UAP, they
are able to show black-box\footnote{Black-box refers to attacks with full access
to the training dataset but no knowledge of the DNN architecture of learned
parameter matrices.} results which are time shift invariant, which is a
limitation of FGSM.  Additionally, the authors use the energy ratios of the 
perturbation and modulated signal as an attack constraint, a metric that the current
work uses as well.  However, the authors consider perturbations which are below
the noise floor but implicitly assume they are able to compromise the
eavesdropper's signal processing chain by not considering the effect of the 
channel on the perturbation signal.  Therefore, \cite{RN100} only considers
attacks that have direct-access to the classifier and aren't transmitted OTA.
The current work expands upon the study of white-box\footnote{White-box refers to
attacks with perfect knowledge of the learned parameter matrices of the DNN.} 
direct-access attacks against RFML systems by exploring the vulnerabilities 
versus neural network input size.  Additionally, the current work considers
white-box self-protect attacks, which are launched OTA, where receiver effects
can negatively impact adversarial success and must also be evaluated against the 
effect the perturbation has on the underlying signal transmission by
characterizing the BER.

% ------------------------------------------------------------------------------
\section{Threat Model for RFML} \label{sec:threat}
% ------------------------------------------------------------------------------
A rich taxonomy already exists for describing threat models for adversarial
machine learning in the context of CV; however, threat models which only 
consider CV applications lack adversarial goals and capabilities that are unique
to RFML.  Therefore, the current work extends the threat model initially 
proposed in \cite{RN13} for RFML in Figure \ref{fig:threat:ThreatModel}.
This section first describes the system model considered for AMC and
then expands on the unique categories of adversarial goals and capabilities
that must be considered when discussing adversarial threats to RFML systems.

% ------------------------------------------------------------------------------
\subsection{Automatic Modulation Classification System Model} \label{sec:threat:model}
The current work considers the task of blind signal classification where an 
eavesdropper attempts to detect a signal in the spectrum, isolate it in time and
frequency, and perform modulation classification.  This task assumes that the
signal is a wireless communication between a transmitter and a cooperative
receiver where the eavesdropper is not synchronized and has very limited
\emph{a priori} information about the communication.  Ultimately, the
eavesdropper could then use the output for DSA, signals intelligence, and/or as 
a preliminary step to demodulating the signal and extracting the actual
information transmitted.

The study of adversarial examples in this model could be framed from the 
perspective of either the eavesdropper or the transmitter.  First, this study
can be considered a vulnerability analysis of RFML systems and the information
gained can then be used to produce a more robust eavesdropper that is hardened
against deception by adversarial machine learning.  Additionally, this study
could be considered a feasibility analysis for methodology to protect
transmissions from eavesdroppers.  Evading an eavesdropper can limit tracking of
the transmitter or automatic demodulation of its transmission.  The current work
does not take a side in the application of this technology and presents a case
for both sides; however, the term adversary is used to describe the transmitter
that seeks to evade an eavesdropper for the remainder of the current work.

% ------------------------------------------------------------------------------
\subsection{Adversarial Goals} \label{sec:threat:goals}
Three main goals are traditionally considered for adversarial machine learning
\cite{RN13}: confidence reduction, untargeted misclassification, and targeted
misclassification.  Confidence reduction is the easiest goal an adversary can
have.  It simply refers to introducing uncertainty into the classifier's
decision even if it ultimately determines the class of signal correctly.  An
adversary whose goal is simply to be classified as any other signal type than
its true class, can be described as untargeted misclassification.
Targeted misclassification is typically the most difficult goal of adversarial
machine learning.  It occurs when an adversary desires a classifier to output a
specific target class instead of simply any class that is not the true class. 
Due to the hierarchical nature of human engineered modulations, the difficulty
of targeted misclassification for AMC depends heavily on the signal formats of
the true and target class.  Targeted misclassification are sometimes split
between attacks that start with a real input \cite{RN68, RN75} versus those that
start with noise \cite{RN11}.  The threat model presented in Figure
\ref{fig:threat:ThreatModel} only considers the former because the current work
assumes that an adversary's primary goal is to transmit information and not
simply degrade classifier performance.

Further, the current work categorizes adversarial goals based on where the 
attack is launched from.

\subsubsection{Direct Access} \label{sec:threat:goals:directAccess}
Traditional adversarial machine learning, such as those generally considered
in CV or the attack considered in \cite{RN100}, fall into the direct access
category.  This category of attack is performed ``at the eavesdropper'' as part
of their signal processing chain.  Therefore, the propagation channel and
receiver effects for the example is known at the time of crafting the
perturbation, the perturbation is not subjected to any receiver effects, and the
perturbation will have no effect on the intended receiver because it is not sent
OTA.  Attacks at this level are very useful for characterizing the worst case
vulnerabilities of a classifier but they are less realistic in the context of
RFML because it assumes that the signal processing chain has been compromised.

\subsubsection{Self Protect} \label{sec:threat:goals:selfProtect}
When the adversarial perturbation is added at the transmitter and propagates
along with the transmitted signal to the eavesdropper, this can be categorized
as self protect.  By adding the perturbation at the transmitter, the 
perturbation can still be completely synchronous with the signal transmission; 
however, the perturbation will now be subjected to all of the receiver effects
traditionally considered in RFML and will also impact the intended receiver. 
While many of the algorithms that are successful for the direct access category
of attacks will be applicable to self protect, the evaluation of adversarial
success must take into account receiver effects.  Therefore, attacks that seek
to create minimal perturbations, such as the modified FGSM method presented in
\cite{RN100}, will no longer work because adversarial success can not be
guaranteed due to the signal being subjected to a stochastic process.

\subsubsection{Cover} \label{sec:threat:goals:cover}
RFML allows for a third category of adversarial goals, in which the adversarial
perturbation originates from a separate emitter from the transmitter and is only
combined at the eavesdropping device.  Low cost transmitters can be size,
weight, and power (SWaP) constrained.  Therefore, it may be beneficial to have
a single unit provide cover for multiple SWaP constrained nodes.  However,
because these attacks cannot rely on synchronization between the transmission
and perturbation, the perturbations must be time shift invariant \cite{RN100}
making this category of attack more difficult.  The current work does not
present a study of this category of adversarial goal and leaves that to future
work.

% ------------------------------------------------------------------------------
\subsection{Adversarial Capabilities} \label{sec:threat:capabilities}
Traditional adversarial machine learning capabilities, such as those described
in \cite{RN13}, generally help with determining ``what you want a classifier to
see'' by providing information about the target DNN that can subsequently be
used to optimize the input.  In the most extreme case, attacks may have perfect
knowledge of the learned parameters of the model.  These attacks are referred to
as white-box.  In a slightly more realistic case, the attacker may have access
to the network architecture and training dataset, but not the learned
parameters.  The attacker must then create adversarial examples that generalize
over all possible models created from the dataset and architecture.  In a very
limited case, the attacker may only have access to what is deemed an oracle, an
entity that will label a limited number of $X, Y$ pairs for the attacker through
an API \cite{RN53} or an observable wireless transmission \cite{RN102, RN44}. 
This allows the attacker to perform limited probes against the target network in
order to build up an attack.

Adversarial machine learning applied to RFML has a different class of
capabilities an attacker can possess that can be thought of as ``the ability to
make a classifier see a specific example''.  RF propagation can be directed 
through the use of smart antennas.  Therefore, if a transmitter knew the
location of the receiver, it could direct its energy only at the receiver, thus
minimizing the signal-to-noise ratio (SNR) at the eavesdropper.  Similarly, a
jammer could direct energy only at the eavesdropper, maximizing the impact of
perturbations on classification accuracy while minimizing the impact to the
receiver.

Signal processing chains can present an impediment to adversarial success. 
Traditionally, RF front ends are built to reject out of band interference and
therefore adversarial perturbations consisting of high frequencies could be 
filtered out.  Power amplifiers can exhibit non-linear characteristics which
would distort the perturbation.  Further, the precision of the analog to digital
converter could limit the attack to stair stepped ranges.

% ------------------------------------------------------------------------------
\subsection{Threat Model Assumed in the Current Work} \label{sec:threat:ourWork}
In the current work we assume direct access to the learned parameters of the
target DNN and set the goal as untargeted misclassification.  The current work
considers perturbations that are specific to the underlying transmitted signal 
and characterizes their effectiveness in the presence of receiver effects such
as noise, sample time offsets, and frequency offsets.  Therefore, both direct
access attacks as well as self protect are considered.  The current work does
not assume knowledge of either the eavesdropper or receiver locations and
therefore does not consider directional antennas and instead shows results
across varying SNR ranges.   Further, the current work assumes that the receiver
is fixed and thus does not introduce any modifications to the receive chain.

% ------------------------------------------------------------------------------
\section{Methodology} \label{sec:methodology}
\begin{figure}
    \includegraphics[width=\linewidth]{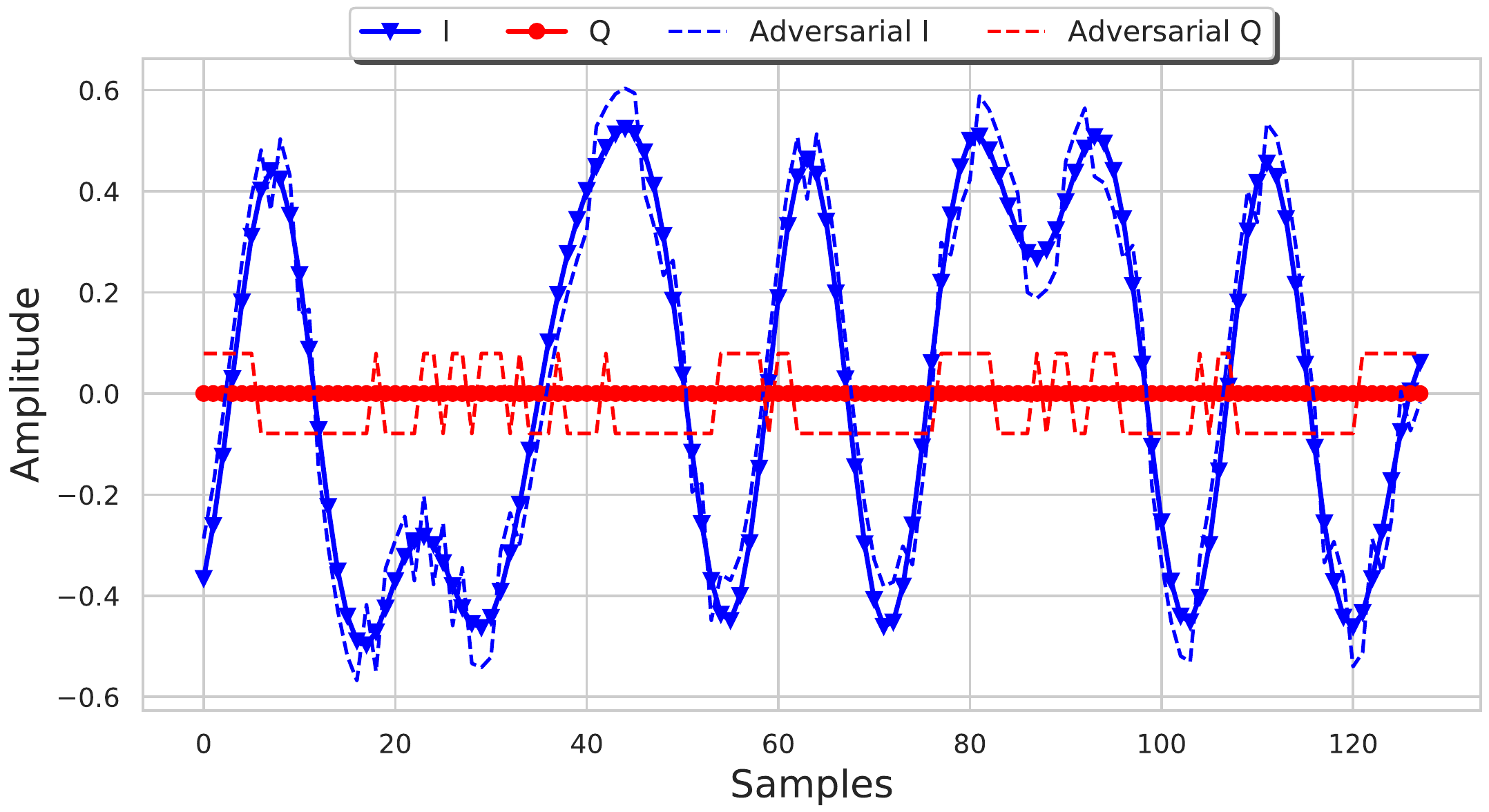}
    \caption{BPSK adversarial example with a 10 dB ($E_s/E_j$) perturbation, created with the FGSM \cite{RN68} algorithm, applied.}
    \label{fig:bpskAdversarial}
\end{figure}

\begin{figure*}
    \includegraphics[width=\linewidth]{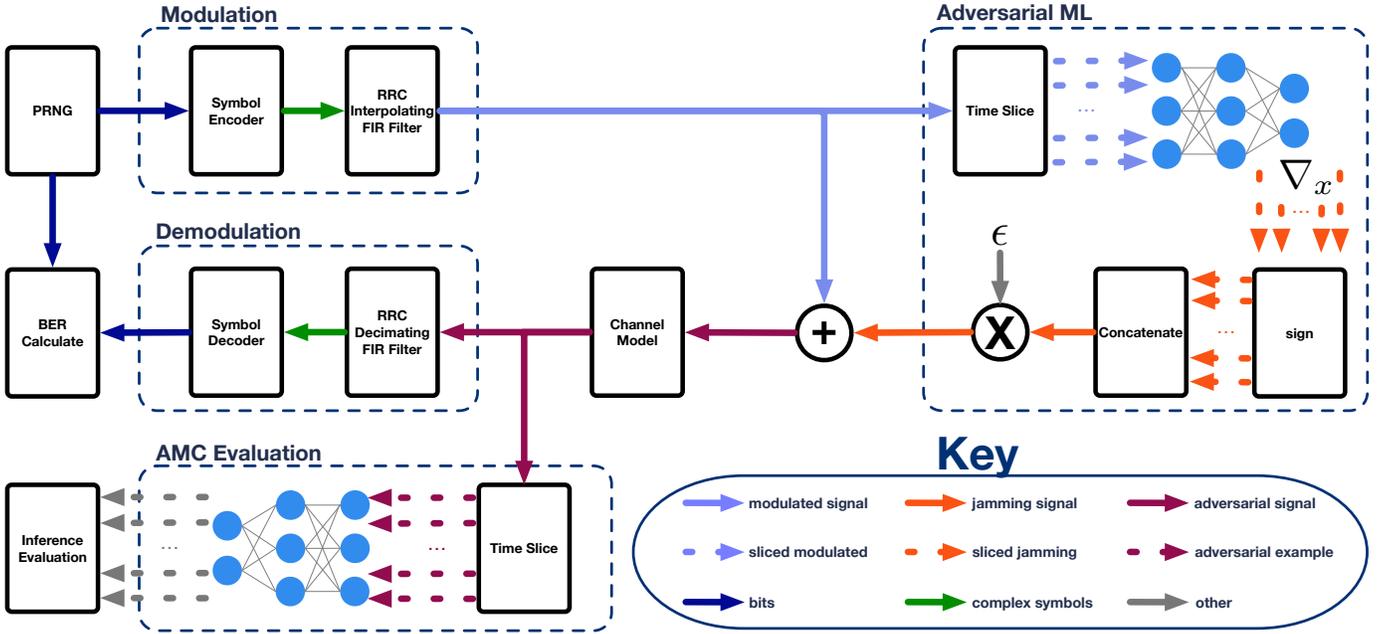}
    \caption{Block diagram of the evaluation methodology developed for the current work.  The current work assumes perfect knowledge of the target DNN and therefore the DNN shown in the AMC Evaluation and Adversarial ML blocks are identical and simply separated for clarity.}
    \label{fig:methodology:ModemFG}
\end{figure*}

Most raw-IQ based signal classifiers seek to take in a signal snapshot, 
$\bm{x}$, and output the most probable class $\bm{y}$.  Traditionally, $\bm{x}$
would represent a single channel of complex samples, with little pre-processing 
performed, and could therefore be represented as a two-dimensional matrix
$[\text{IQ}, \text{number of samples}]$.  Specifically, RFML systems, which
generally use DNNs, learn a mapping from the data by solving

\begin{equation}
	\argmin_{\bm{\theta}} \mathcal{L}(f(\bm{\theta}, \bm{x}), \bm{y})),
	\label{eq:ml}
\end{equation}

where $\bm{x}$ and $\bm{y}$ represent the training inputs and target labels
respectively and $f$ represents the chosen network architecture.  A loss 
function ($\mathcal{L}$), such as categorical cross entropy, is generally used
in conjunction with an optimizer, such as stochastic gradient descent or Adam
\cite{RN108}, to train the DNN and thus learn the network parameters
$\bm{\theta}$. While training the model, the dataset is fixed (assuming no data
augmentation) and is assumed to be sampled from the same distribution that will
be seen during operation of the RFML system.

Untargeted adversarial machine learning is simply the inverse of this process.
By seeking to maximize the same loss function, an adversary can decrease the
accuracy of a system.  Therefore, the adversary is also solving an
optimization problem that can be defined by the following.

\begin{equation}
	\argmax_{\bm{x}^*} \mathcal{L}(f(\bm{\theta}, \bm{x}^*), \bm{y}))
	\label{eq:adversarialml}
\end{equation}

In this case, the parameters, $\bm{\theta}$, of the classifier are fixed but the
input, $\bm{x}^*$, can be manipulated.  Many approaches exist to solve this
problem.  In particular, FGSM \cite{RN68} creates untargeted adversarial
examples using

\begin{equation}
	\bm{x}^* = \bm{x} + \epsilon \times \operatorname{sign}(\nabla_x\mathcal{L}(f(\bm{\theta}, \bm{x}), \bm{y})),
	\label{eq:fgsm}
\end{equation}

where $\bm{y}$ represents the true input label and $\nabla_x$ represents the
gradient of the loss function with respect to the original input, $\bm{x}$.
This methodology creates adversarial examples constrained by a distance,
$\epsilon$, in the feature space in a single step.  $\bm{x}^*$ is referred to as
an adversarial example.  One adversarial example used in the current work is
presented in Figure \ref{fig:bpskAdversarial}, where the source modulation is
BPSK and a perturbation has been applied to achieve untargeted evasion for a
direct access attack.

In the context of wireless communications, the absolute value of the signal is
generally less important than the relative power of the signal with respect to
some other signal such as noise.  Therefore, similar to \cite{RN100}, the 
current work reformulates the perturbation constraint, $\epsilon$, from a
distance bounding in the feature space to a bounding of power ratios. 
Additionally, the signal can be directly evaluated on the primary metric of
interest, BER, as opposed to the use of human perception, or a proxy for it such
as $\epsilon$, in CV.  Further, 

% ------------------------------------------------------------------------------
\subsection{Adapting FGSM} \label{sec:methodology:fgsm}
The average energy per symbol ($E_s$) of a transmission can be computed using

\begin{equation}
    \E[E_{s}] = \frac{\text{sps}}{N} \sum_{i=0}^{N} |s_i|^2,
	\label{eq:avgEnergyPerSymbol}
\end{equation}

where $\text{sps}$ represents samples per symbol, $N$ is the total number of
samples, and $s_i$ represents a particular sample in time.  Without loss of
generality, the current work assumes the average energy per symbol of the 
modulated signal, $E_s$, is $1$.  Therefore, the power ratio of the underlying
transmission to the jamming/perturbation signal\footnote{Because the 
perturbation is an electronic signal deliberately crafted to impair the
successful operation of the eavesdropper, the current work uses jamming signal
and perturbation signal interchangeably.} ($E_j$) can be derived as

\begin{equation}
    \begin{aligned}
        \frac{E_s}{E_j} & = \frac{1}{E_j}\\
            & = 10^{-E_j \text{(dB)} / 10}
    \end{aligned}
    \label{eq:EsEj}
\end{equation}

Since the input of $\operatorname{sign}(\nabla_x)$ in (\ref{eq:fgsm}) is 
complex, the output is also complex, and is therefore a vector whose values
are $(\pm 1, \pm 1j)$. Therefore, the magnitude of each sample of the jamming 
signal can be computed as

\begin{equation}
    \begin{aligned}
        |\operatorname{sign}(\nabla_x)| & = |\operatorname{sign}(z)|\\
            & = \sqrt{(\pm 1)^2 + (\pm 1)^2}\\
            & = \sqrt{2}\\
    \end{aligned}
    \label{eq:signedGradientMag}
\end{equation}

Thus the energy per symbol of $\operatorname{sign}(\nabla_x)$ can be computed
by plugging (\ref{eq:signedGradientMag}) into (\ref{eq:avgEnergyPerSymbol})
resulting in

\begin{equation}
    \begin{aligned}
            E_{\operatorname{sign}(\nabla_x)} & = \frac{\text{sps}}{N} \sum_{i=0}^{N} |\operatorname{sign}(\nabla_x)|^2\\
                         & = 2 \times \text{sps}\\
    \end{aligned}
    \label{eq:EjOriginal}
\end{equation}

Because $\text{sps}$ is fixed throughout transmission, a closed form scaling 
factor, $\epsilon$, can be derived to achieve the desired energy ratio
($E_s/E_j$) by using

\begin{equation}
    \begin{aligned}
        \epsilon & = \sqrt{\frac{\frac{E_s}{E_j}}{E_{\operatorname{sign}(\nabla_x)}}}\\
             & = \sqrt{\frac{10^\frac{-E_j}{10}}{2 \times \text{sps}}}
    \end{aligned}
    \label{eq:scalingJ}
\end{equation}

Plugging $\epsilon$ into (\ref{eq:fgsm}) allows the creation of adversarial
examples constrained by $E_s/E_j$ and can be succinctly defined as

\begin{equation}
	\bm{x}^* = \bm{x} + \sqrt{\frac{10^\frac{-E_j}{10}}{2 \times \text{sps}}} \times \operatorname{sign}(\nabla_x\mathcal{L}(f(\bm{\theta}, \bm{x}), \bm{y}))
    \label{eq:powerConstrainedFGSM}
\end{equation}

Constraining the power ratio in this way can be useful for evaluating system
design trade-offs.  Any given transmitter has a fixed power budget and the
current work considers an adversarial machine learning technique which is not
aware of the underlying signal; therefore, power which is used for the jamming 
signal subsequently cannot be used for the underlying transmission.  Future 
adversarial machine learning techniques could take into account the bit error
rate in their methodology which would allow for this energy to accomplish both
purposes, but, this exploration is left to future work.

% ------------------------------------------------------------------------------
\subsection{Simulation Environment} \label{sec:methodology:simulation}
The high level overview of the simulation environment used in the current work
is shown in Figure \ref{fig:methodology:ModemFG} and each major block is 
described below.  Full evaluation in the context of wireless communications
requires the interfacing of both a DSP and ML framework.  The current work uses
GNU Radio and PyTorch respectively; however, the methodology is not dependent
upon use of those frameworks in any way.

\subsubsection{Modulation} \label{sec:methodology:simulation:modulation}
The initial modulated signal is generated by a simple flow graph in GNU Radio.
Unless otherwise stated, the parameters for transmission can be summarized as
follows.  The symbol constellations used are BPSK, QPSK, 8PSK, and QAM16.
The root raised cosine filter interpolates to $8$ samples per symbol using a
filter span of $8$ symbols and a roll-off factor of $0.35$.  $1000$ examples per
modulation scheme are created using a random bit stream.

\subsubsection{Adversarial ML} \label{sec:methodology:simulation:adversarial}
In order to craft the jamming signal using adversarial machine learning
techniques it is necessary to first slice the signal into discrete examples
matching the DNN input size.  Before feeding these examples into the DNN,
dithering is employed to add small amounts of noise to the examples.  The FGSM
algorithm is then used to create the perturbations which are concatenated back
together to form the jamming signal.  For each $E_s/E_j$ studied, the
jamming signal is scaled linearly using (\ref{eq:scalingJ}) and added to the
modulated signal.  Unless otherwise stated, $E_s/E_j$ is swept from $0$ to $20$
dB with a step size of $4$ dB.

\subsubsection{Channel Model} \label{sec:methodology:simulation:channel}
The current work considers a simple channel model with Additive White Gaussian 
Noise (AWGN) and center frequency offsets.  The received signal can be
characterized as follows:

\begin{equation}
    \begin{aligned}
        S_{\text{rx}}(t) &= e^{-j 2 \pi f_{\text{o}} t} S_{\text{tx}}(t) +  \mathcal{N}(0,\,\sigma^{2})
    \end{aligned}
    \label{eq:methodology:channel}
\end{equation}

Where $f_{o}$ is the normalized frequency offset and $\sigma^2$ is given
by the desired $E_s/N_0$.  The channel model is again implemented using a
GNU Radio flow graph.

\subsubsection{Demodulation} \label{sec:methodology:simulation:demodulation}
Demodulating the received signal consists of match filtering, down-sampling to
one sample per symbol, and decoding the symbols back into a bit stream to verify
the data received matches the data transmitted.  The demodulation is also 
implemented as a GNU Radio flow graph and assumes both symbol and frame
synchronization.

\subsubsection{Automatic Modulation Classification Evaluation} \label{sec:methodology:simulation:classification}
Top-1 accuracy is the metric used for classifier evaluation in \cite{RN27}, 
\cite{RN28}, and\cite{RN12} and is the metric we use for evaluation in the 
current work.  For untargeted adversarial machine learning, adversarial success
is defined as a lower Top-1 accuracy as opposed to a higher accuracy.

% ------------------------------------------------------------------------------
\subsection{Automatic Modulation Classification Target Network} \label{sec:methodology:amc}
% ------------------------------------------------------------------------------
\subsubsection{Network Architecture} \label{sec:methodology:amc:architecture}
The current work uses the DNN architecture first introduced in \cite{RN27} for
raw-IQ modulation classification.  This architecture consists of two 
convolutional layers followed by two fully connected layers.  This network takes
the IQ samples as a $[1, 2, N]$ tensor which corresponds to 1 channel, IQ, and N
input samples.  The current work uses extended filter sizes as done in
\cite{RN28} and \cite{RN12}, using filters with 7 taps and padded with 3 zeros
on either side.  The first convolutional layer has 256 channels, or kernels, and
filters I and Q separately.  The first layer does not use a bias term as this
led to vanishing gradients during our training.  The second layer consists of 80
channels and filters the I and Q samples together using a two-dimensional real
convolution.  This layer includes a bias term.  The feature maps are then
flattened and fed into two fully connected layers, the first consisting of 256
neurons and the second consisting of the number of output classes.  All layers
use ReLU as the activation function (except for the output layer).  As a
pre-processing step, the average power of each input is normalized to $1$.

% ------------------------------------------------------------------------------
\subsubsection{Dataset A} \label{sec:methodology:amc:datasets:rml}
The majority of this work uses the open source RML2016.10A dataset introduced in
\cite{RN29}.  This synthetic dataset consists of 11 modulation types: BPSK,
QPSK, 8PSK, CPFSK, GFSK, PAM4, QAM16, QAM64, AM-SSB, AM-DSB, and WBFM.  These 
signals are created inside of GNU Radio and passed through a dynamic channel 
model to create sample signals at SNRs ranging from -20dB to 18dB.

Using an open source dataset allows for quick comparison of results to those 
seen in literature; however, this dataset only provides one input size, $128$
complex samples.  Furthermore, this dataset contains limited center frequency
offsets.  Therefore, it was necessary to create an additional dataset to perform
all of the evaluations contained in the current work.

% ------------------------------------------------------------------------------
\subsubsection{Dataset B} \label{sec:methodology:amc:datasets:custom}
This additional dataset was also created using synthetic data from GNU Radio.
Three datasets were created with varying input size (128, 256, and 512).
These synthetic datasets consists of $5$ modulation schemes: BPSK, QPSK, 8PSK,
QAM16, and QAM64.  Keeping with the RML2016.10A Dataset, the samples per symbol
of the root raised cosine filter were fixed at $8$.  The one sided filter span
in symbols is varied uniformly from 7 to 10 with a step size of 1.  The roll-off
factor of the root raised cosine was varied uniformly from $0.34$ to $0.36$ with
a step size of $0.01$.  For the channel model, the modulated signal was 
subjected to AWGN and given a center frequency offset as described by
(\ref{eq:methodology:channel}) to simulate errors in the receiver's signal
detection stage \cite{RN12} .  The power of the AWGN is calculated using
$E_s/N_o$ and varied uniformly from $0$ dB to $20$ dB with a step size of $2$. 
The center frequency offset, which was normalized to the sample rate, is swept
uniformly from $-1\%$ to $1\%$ with a step size of $0.2\%$.

% ------------------------------------------------------------------------------
\subsubsection{Training Results} \label{sec:methodology:amc:results}
\begin{figure}
    \includegraphics[width=\linewidth]{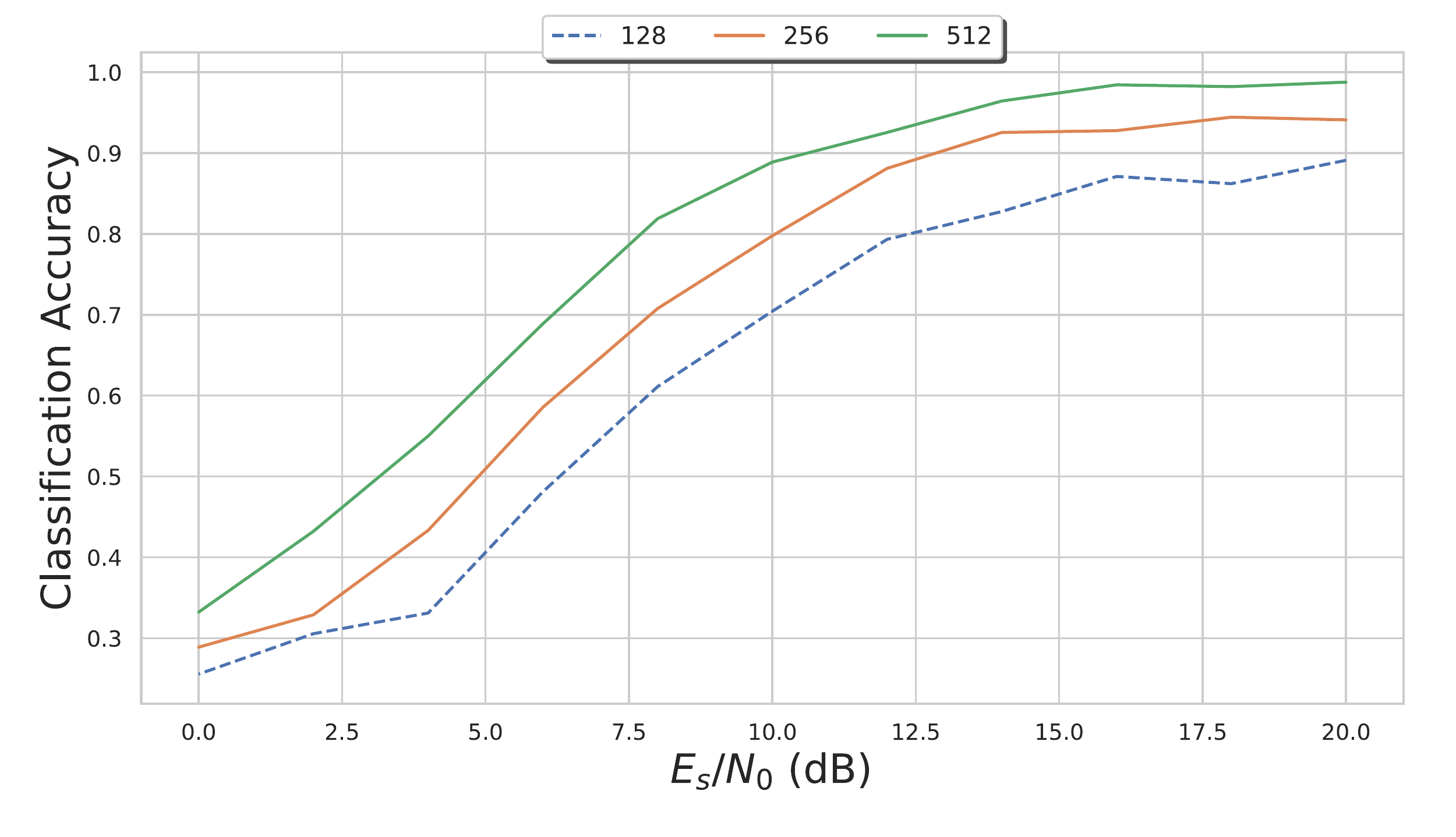}
    \caption{Dataset B test accuracy vs SNR for three different neural network input sizes.  As expected, increasing the input size results in increasing test accuracy over the entire SNR range studied.}
    \label{fig:amc:accVsSnrCustom}
\end{figure}

The network is implemented in PyTorch and trained using an NVIDIA 1080 GPU with
the Adam \cite{RN108} optimizer.  The batch size used is 1024 when the network 
is trained with Dataset A and 512 when trained with Dataset B due to the
increased example sizes.  Models trained on Dataset A use dropout for
regularization, as was initially proposed in \cite{RN27}; however, models
trained on Dataset B use Batch Normalization as this increased training
stability for the larger example sizes.  For all models, the learning rate is
set to $0.001$ and early stopping is employed with a patience of $5$.

During training, 30\% of the dataset was withheld as a test set.  The remaining
70\% of the data is used in the training sequence with 5\% of the training set
used as a validation set.  All data is split randomly with the exception that
modulation classes and SNR are kept balanced for all sets.  Each of the
models is then evaluated at each SNR in the test set for overall accuracy and
the results are shown in Figure \ref{fig:amc:accVsSnrCustom}.  As expected, 
increasing the input size lead to increasing accuracy.

% ------------------------------------------------------------------------------
\section{Analysis of Direct Access Attacks} \label{sec:performance}
% ------------------------------------------------------------------------------
%\subsection{Direct Access Attacks} \label{sec:performance:noChannel}
In order to first characterize the effectiveness of adversarial machine learning
on raw-IQ based AMC, a baseline study of average classification accuracy against
$E_s/E_j$ was performed using the model trained on Dataset A.  This attack was
performed with no noise added to the adversarial examples and thus assumes
direct access to the classifier input.

\begin{figure}
    \includegraphics[width=\linewidth]{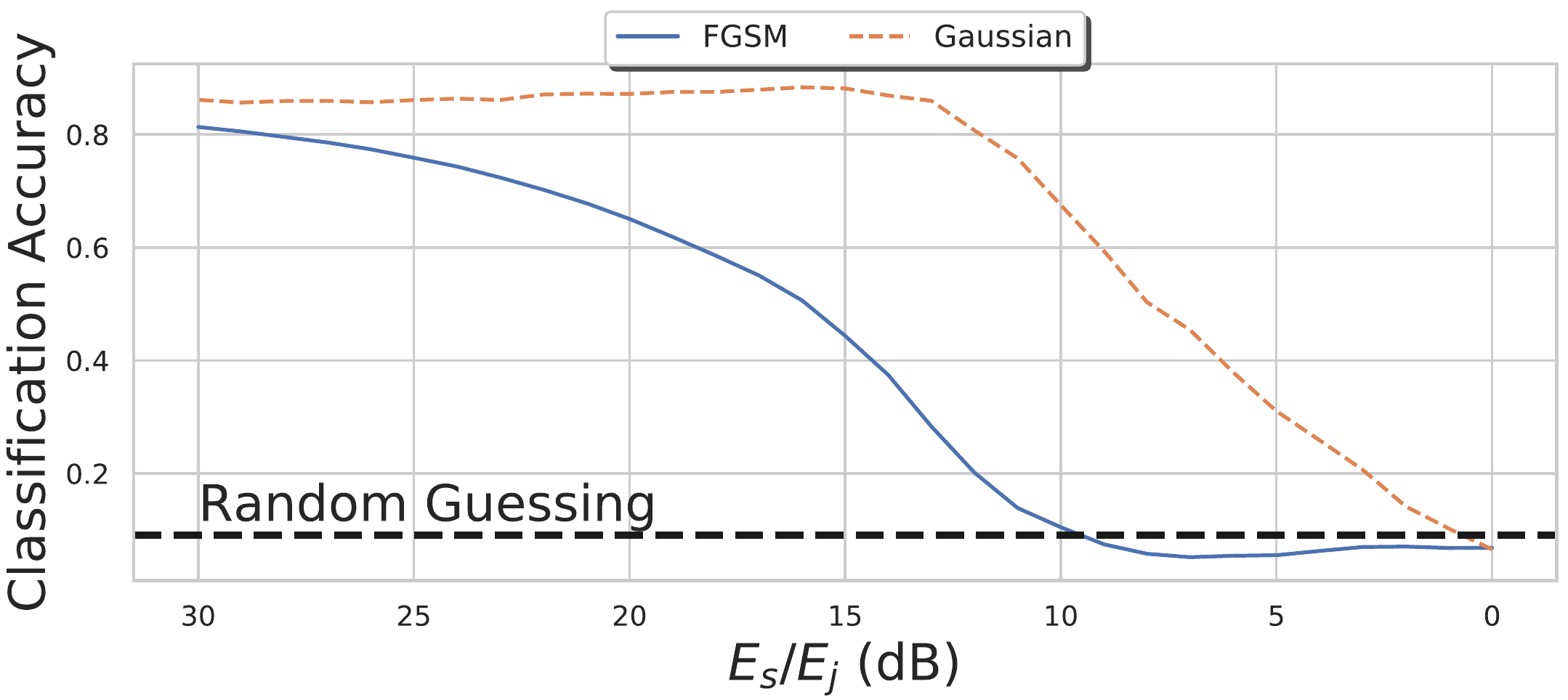}
    \vspace{3.5mm}
    \caption{Overall classification accuracy of a model trained on Dataset A for a direct access attack.  This plot compares the classification accuracy when FGSM in used to apply a specific adversarial perturbation to the accuracy when ``jammed'' with a Gaussian noise signal at the same power ratio.}
    \label{fig:performance:averageAccuracyRML}
\end{figure}

As can be seen in Figure \ref{fig:performance:averageAccuracyRML}, even at $30$
dB, the FGSM attack is more effective than simply adding Gaussian noise (AWGN).
At $10$ dB, the FGSM attack is effective enough to degrade the classifier below 
the performance of random guessing.  This represents an $8$ dB improvement over 
the same degradation using Gaussian noise.

For comparison to other results in CV literature, we can plug $E_s/E_j=10$ dB
into (\ref{eq:scalingJ}) which yields that an $\epsilon$ of $\approx 0.079$ is
sufficient for accomplishing the goal of untargeted adversarial machine learning
for direct access attacks on this model.  While this clearly shows an 
improvement over Gaussian jamming, this perturbation is larger than the original
example shown in \cite{RN68} of $0.007$ for performing an untargeted attack
using a source image of a panda.  However, that result used ImageNet as a source
class and GoogLeNet \cite{GoogleNet} as the model where the input dimensions of 
the image were at least $3 \times 256 \times 256$ ($\gg \mathbb{R}^{196,608}$)
while the input size considered here is $1 \times 2 \times 128$
($\mathbb{R}^{256}$).  Therefore, while we know that the underlying
classification task is vastly different and exact perturbation constraints
cannot be directly compared, we next investigate whether increased input
dimensionality makes the model more susceptible to adversarial examples.

% ------------------------------------------------------------------------------
\subsection{Attack Effectiveness versus NN Input Size} \label{sec:performance:inputSize}
\begin{figure}
    \includegraphics[width=\linewidth]{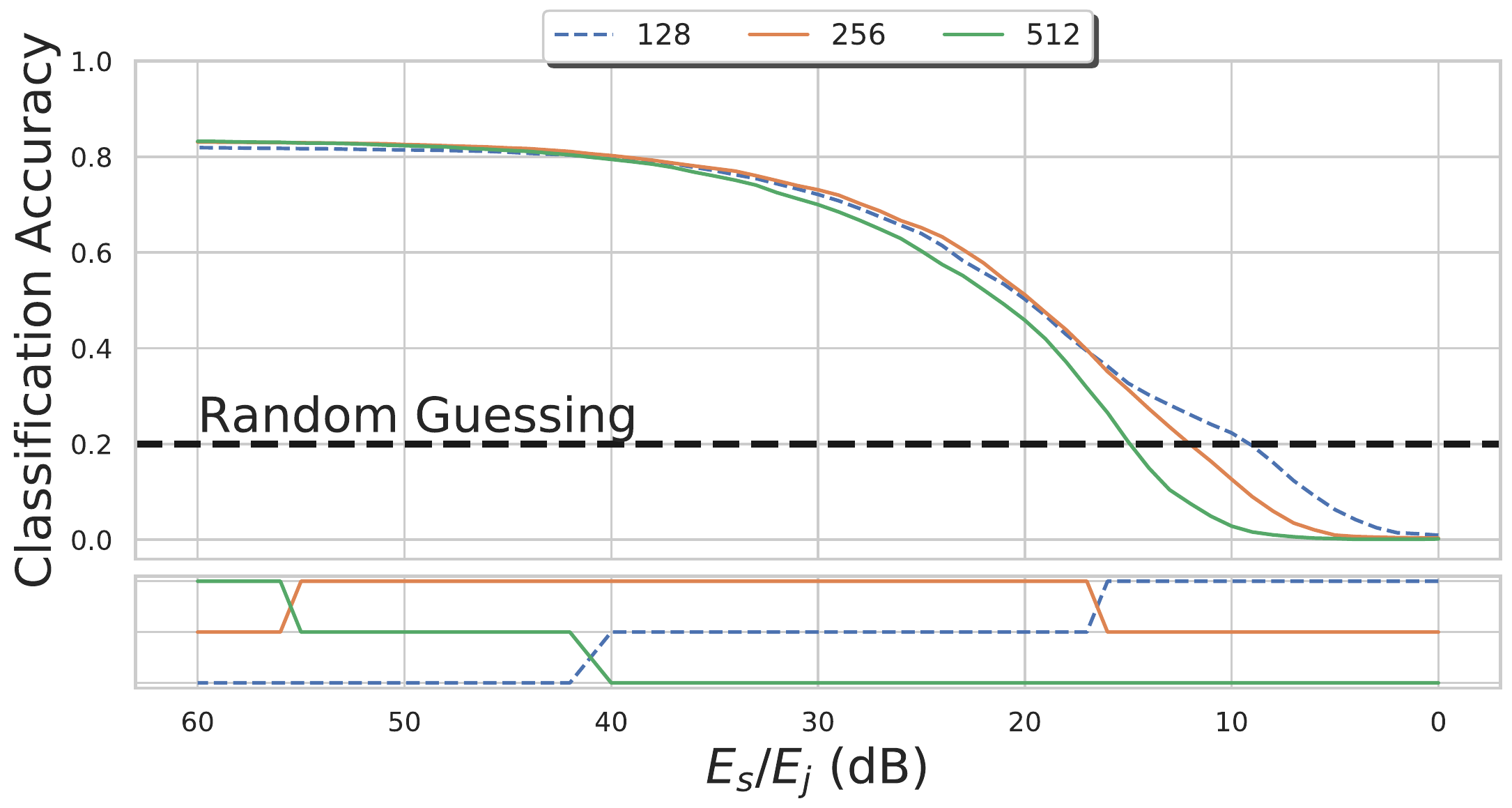}
    \caption{Overall classification accuracy (top) of models trained on Dataset B in the presence of a direct access FGSM attack.  The relative classification accuracy ranking of the three different models for each $E_s/E_j$ (bottom).}
    \label{fig:performance:averageAccuracy}
\end{figure}

Increasing the DNN input size has been empirically shown to improve
the performance of raw-IQ AMC in \cite{RN12} as well as the current work's reproduction
of similar results in Figure \ref{fig:amc:accVsSnrCustom}.  While it is
intuitive that viewing longer time windows of a signal will allow for higher
classification accuracy, it is also intuitive that allowing more adversarial
jamming energy to enter the algorithm will have adverse effects. Therefore, the current
work presents an experiment used to verify this intuition.  Three copies of the
same network, that differ only in input size, are trained on Dataset B.  The
analysis from the previous section is then repeated and shown in Figure
\ref{fig:performance:averageAccuracy}.

As expected, at very high $E_s/E_j$, where the adversarial energy is low, the
network with the largest input size is the most accurate.  However, it is
quickly supplanted by the second largest input size when $E_s/E_j$ drops below
$55$ dB ($\epsilon \approx 0.00044$).  Once $E_s/E_j$ drops below $15$ dB, the
classification accuracy ranking inverts from the initial rankings, with the
smallest input size being the most accurate and the largest input size being the
least accurate.  Therefore, when developing a RFML system for use in adversarial
environments, the benefits of increasing input size must be balanced against the
cost of increasing the attack surface.

% ------------------------------------------------------------------------------
\subsection{Analyzing Individual Adversarial Examples} \label{sec:performance:signalEffect}
While the earlier subsections presented macro-level results, this subsection
presents results at a micro-level by analyzing the fine grained effect of the
adversarial machine learning method on individual examples rather than the
average effect across multiple examples.  The current work considers a single
machine learning example from each of the source modulations\footnote{While
random individual examples are analyzed for simplicity, the conclusions drawn
are further explored in Section \ref{sec:receiver}.}.  For each example,
$E_s/E_j$ is swept from $40$ to $0$ dB with a step size of a $1$ dB.  At each
$E_s/E_j$, the outputs of the DNN before the $\operatorname{softmax}$
function (as was shown in \cite{RN68}) are captured.

One adversarial example for BPSK is shown in Figure \ref{fig:bpskAdversarial}.
It can be seen in the $Q$ samples that, due to the $\operatorname{sign}$
operation in (\ref{eq:powerConstrainedFGSM}), the perturbation applied to the 
signal has a box shape.  Therefore, the perturbation alone is easily
identifiable; however, in the $I$ samples, where the underlying modulated signal
also lies, it is less distinguishable.  Notably, the differences are most
apparent around the symbol locations (note that this signal has $8$ samples per
symbol), which could indicate that the classifier has learned some notion of
synchronization.

% ------------------------------------------------------------------------------
\subsubsection{Difference in Logits} \label{sec:performance:signalEffect:logits}
While the full output of the DNN provides ample information, it is 
multi-dimensional and therefore hard to visualize.  One metric that is often
used is a confusion matrix, which captures the relationships among classes. 
However, confusion matrices are generally only presented as an average across
multiple examples and do not provide any notion of the confidence with which a
classifier made the prediction.  Therefore, a confusion matrix would not
fully capture the variance of the DNN because the outputs would not change
unless the input examples were moved across a decision boundary.  Another
metric that could be used is to apply the $\operatorname{softmax}$ function to
the output and report the confidence associated with the source class.  This
metric shows the variance of the classifier output but does not provide any
indication of the Top-$1$ accuracy score because even a low confidence output
could still be the highest and therefore the predicted class.

The current work presents an additional metric, which we term the ``difference 
in logits'' ($\Delta_{logits}$), that simultaneously captures the accuracy of 
the classifier as well as the variance in outputs.  ``Logits'' refers to the
DNN output before the $\operatorname{softmax}$ function has been applied.
The maximum output of all incorrect classes is subtracted from the source
(true) class output, which can be described by the following Equation.

\begin{equation}
    \Delta_{logits} = y_s - \max(y_i \forall{i \neq s})
\end{equation}

The difference in logits can be visualized as the shaded region in the top of
Figures \ref{fig:performance:bpsk} and \ref{fig:performance:qam16}.  When
$\Delta_{logits}$ is positive, the example is correctly classified and a negative
$\Delta_{logits}$ therefore indicates untargeted adversarial success.

% ------------------------------------------------------------------------------
\subsubsection{Classifier Output versus $E_s/E_j$} \label{sec:performance:signalEffect:sweep}
\begin{figure}
    \includegraphics[width=\linewidth]{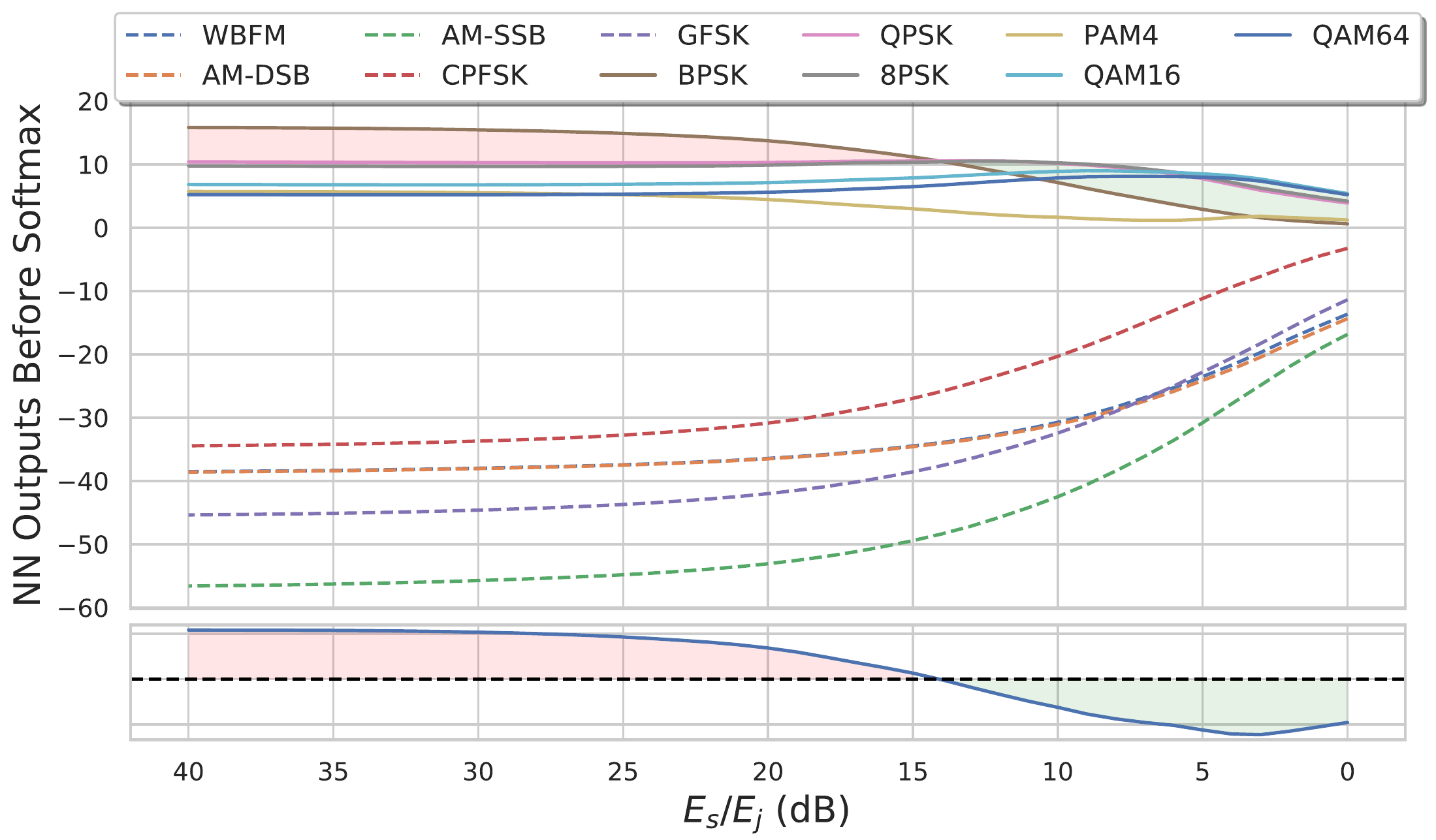}
    \caption{Output of the model trained on Dataset A for a direct access FGSM attack using a single BPSK adversarial example across varying $E_s/E_j$ (top) and the corresponding difference in logits (bottom).}
    \label{fig:performance:bpsk}
\end{figure}

The output of the classifier for the BPSK example, across multiple $E_s/E_j$ is 
shown in Figure \ref{fig:performance:bpsk}.  At an $E_s/E_j$ of $10$ dB, the
jamming intensity present in Figure \ref{fig:bpskAdversarial}, untargeted
misclassification is achieved because the BPSK output is not the highest output
of the classifier; this result is also indicated by viewing that 
$\Delta_{logits}$ is negative.  However, even though misclassification is 
achieved, the signal is still classified as a linearly modulated signal, with
the predicted modulation order increasing as $E_s/E_j$ increased.  Linearly
modulated signals have symbols which exist in the IQ plane (distinguished as
solid lines in Figure \ref{fig:performance:bpsk}) versus a FSK or continuous 
signal (distinguished as dashed lines) whose symbols exist in the frequency
domain or do not have discrete symbols at all, respectively.  Therefore, while
the adversarial machine learning method was able to achieve untargeted
misclassification by causing the classifier to misinterpret the specific
linearly modulated signal, the classifier still captured the hierarchical family
of the human-engineered modulation.  This reinforces the natural notion that the
difficulty of targeted adversarial machine learning varies based on the specific
source and target modulations used.

\begin{figure}
    \includegraphics[width=\linewidth]{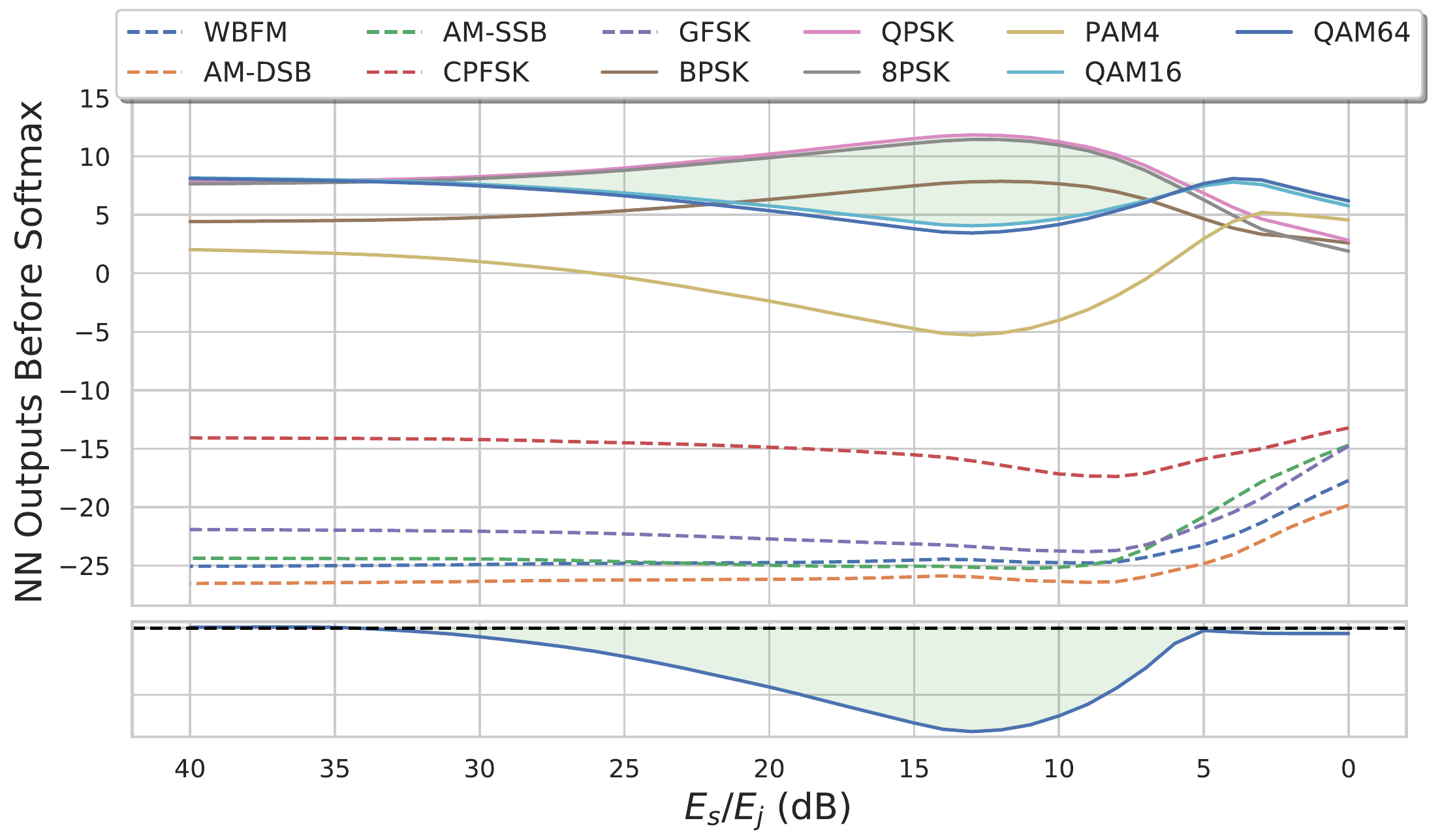}
    \caption{Output of the model trained on Dataset A for a direct access FGSM attack using a single QAM16 adversarial example across varying $E_s/E_j$ (top) and the corresponding difference in logits (bottom)}
    \label{fig:performance:qam16}
\end{figure}

Figure \ref{fig:performance:qam16} shows the output of the classifier for a
single QAM16 example.  As was observed in Figure \ref{fig:performance:bpsk},
at very low $E_s/E_j$, where the attack intensity is the highest, the example
is again classified as QAM (though untargeted misclassification is narrowly
achieved because the model believes it is QAM64).  Further, the QAM16 example
required much lower energy ($E_s/E_j < 30$ dB) than the BPSK example
($E_s/E_j < 15$ dB) to achieve untargeted misclassification.  Therefore,
increasing the perturbation energy does not always provide advantageous effects
from the evasion perspective, as can be observed from the difference in logits 
of Figure \ref{fig:performance:qam16}, and the optimal attack intensity varies
between source modulations.

% ------------------------------------------------------------------------------
\subsubsection{Mutation Testing with AWGN} \label{sec:performance:signalEffect:awgn}
Mutation testing was proposed as a defense in \cite{RN38} where the authors
repeatedly applied domain specific noise to a machine learning example and
calculated the input's sensitivity, with respect to the classifier output, in
the presence of this noise.  The authors of \cite{RN38} found that adversarial 
examples were more sensitive to noise than examples contained in the initial
training distribution and therefore mutation testing could be used to detect
adversarial examples.

The current work presents a study of the effect of
AWGN, one of the most prevalent models of noise in RFML, on individual
adversarial examples.  For each $E_s/E_j$, AWGN is introduced to the signal at
varying $E_s/N_0$ (SNR).  $E_s/N_0$ is swept from $20$ to $0$ dB with a step
size of $1$ dB.  For each of the SNRs considered, $1000$ trials are performed.
While $E_s/E_j$ and $E_s/N_0$ are the parameters swept in this experiment, the
jamming to noise ratio ($E_j/N_0$) can be quickly inferred by

\begin{equation}
    \begin{aligned}
        \frac{E_j}{N_0} &= \frac{E_s/N_0}{E_s/E_j}\\
                        &= \frac{E_s}{N_0} \text{dB} - \frac{E_s}{E_j} \text{dB}
    \end{aligned}
    \label{eq:performance:derivedJammingToNoise}
\end{equation}

\begin{figure}[t]
    \includegraphics[width=\linewidth]{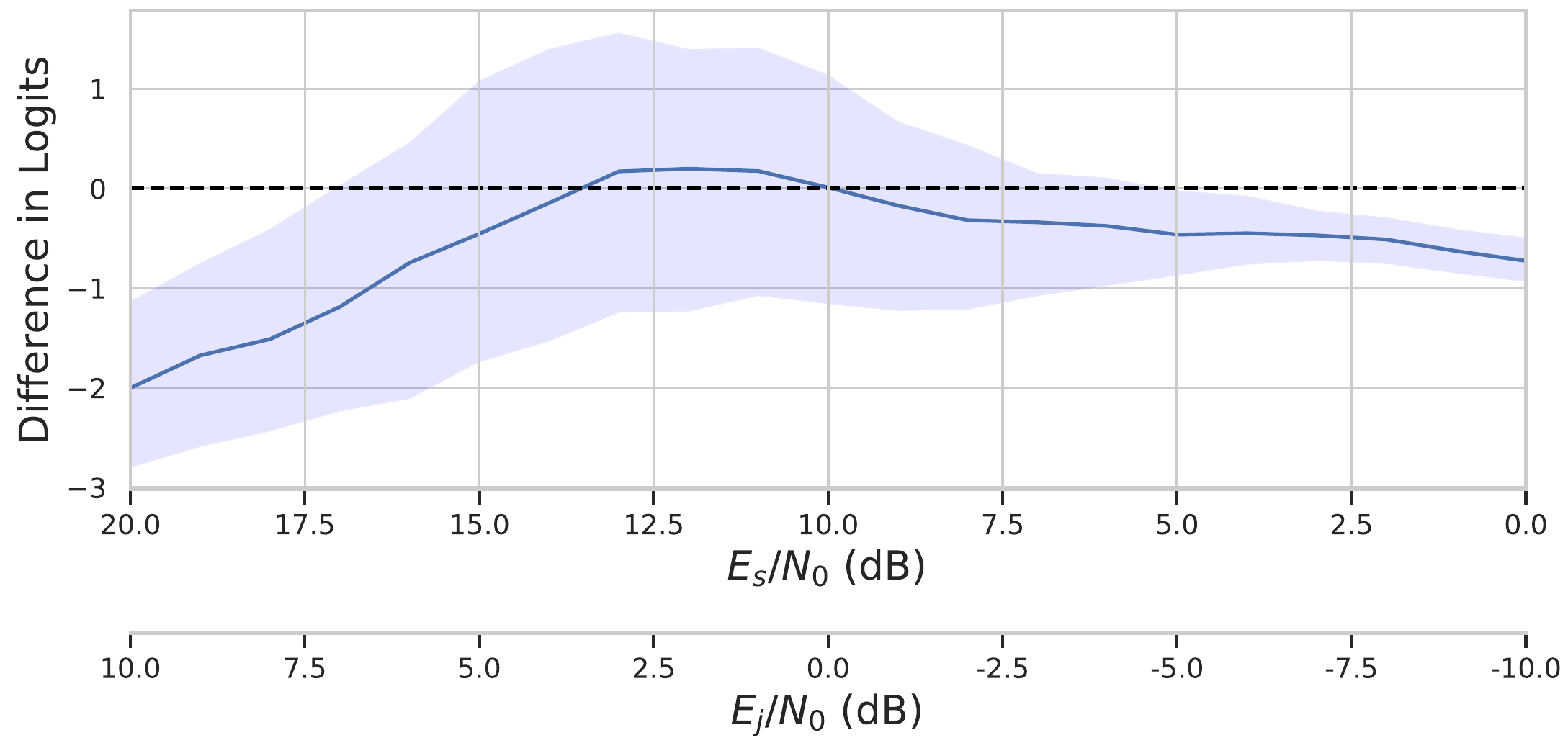}
    \caption{The effect of noise on the output of the model trained on Dataset A for a single BPSK adversarial example with an $E_s/E_j$ of 10 dB.  The line represents the mean of the difference in logits, at a specific $E_s/N_0$, while the shaded region represents the 25th and 75th percentiles.}
    \label{fig:performance:variance:bpsk}
\end{figure}

Again, results are presented in Figure \ref{fig:performance:variance:bpsk} from
the BPSK example originally shown in Figure \ref{fig:bpskAdversarial}, where
$E_s/E_j$ is $10$ dB.  The mean of the difference in logits is shown with the 
25th and 75th percentiles shaded to show the variance in the output of the
classifier at each SNR.  With even a small amount of noise ($E_s/N_0$ of $17$
dB) the 75th percentile of the difference in logits becomes positive indicating
that the example was classified correctly in some iterations.  Increasing the
noise power to roughly half that of the applied perturbation ($E_j/N_0$ of $3$
dB) results in the classification, on average, being correct.

\begin{figure}[t]
    \includegraphics[width=\linewidth]{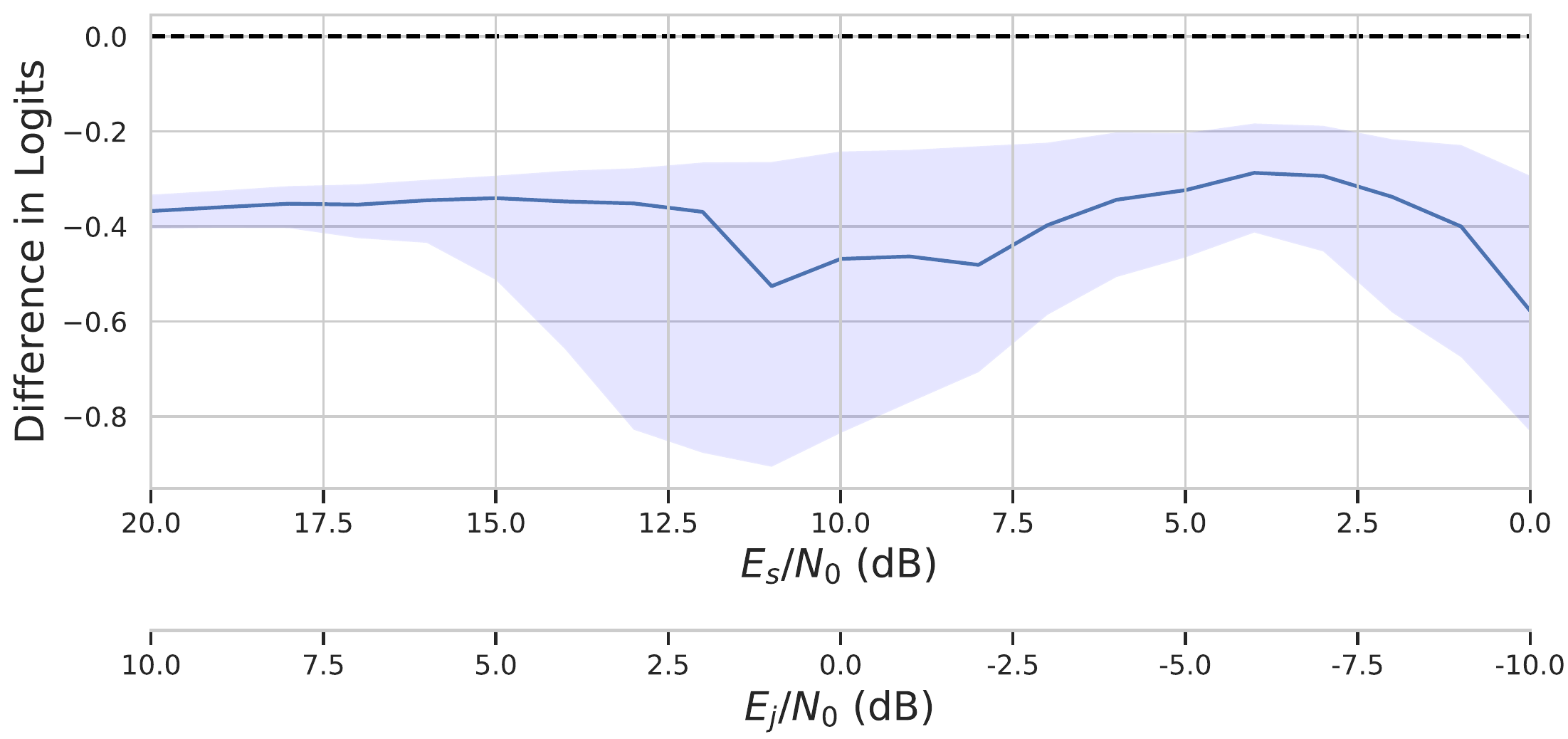}
    \caption{The effect of noise on the output of the model trained on Dataset A for a single QPSK adversarial example with an $E_s/E_j$ of 10 dB.  The line represents the mean of the difference in logits, at a specific $E_s/N_0$, while the shaded region represents the 25th and 75th percentiles.}
    \label{fig:performance:variance:qpsk}
\end{figure}

This effect was not observed across all adversarial examples tested.  In Figure
\ref{fig:performance:variance:qpsk} it is shown that, while the increased
sensitivity of the classifier output is observed in the same range of $E_j/N_0$,
it does not result in a correct classification.  Therefore, while \cite{RN38}
presented general conclusions that all adversarial examples were sensitive to
noise, these results show that this effect is most pronounced when the
adversarial perturbation and noise have similar power.   Additionally, these effects were not observed
at all in the individual 8PSK and QAM16 examples studied. 

This section has shown a baseline result that deep learning based raw-IQ
automatic modulation classification is vulnerable to untargeted adversarial
examples.  Further, it was shown that although increasing the neural network
input size can improve accuracy in non-adversarial scenarios, it can make a
classifier more susceptible to deception for a given $E_s/E_j$.  This section also
showed that noise can have a negative impact on adversarial success.  Therefore,
attacks which can only provide a stochastic input to the classifier (self
protect) must be evaluated differently than attacks that are able to provide a
deterministic input to the classifier (direct access) and thus the following
section presents a more detailed study of self protect attacks using the same
adversarial machine learning method.

% ------------------------------------------------------------------------------
\section{Analysis of Self Protect Attacks} \label{sec:receiver}
% ------------------------------------------------------------------------------
All OTA attacks must consider the impact of receiver effects on adversarial
success; furthermore, self protect attacks must balance the secondary goal of
evading an adversary with the primary goal of transmitting information across a
wireless channel.  Neither of these effects have been considered in prior work
and therefore, while the previous section studied adversarial success in near
perfect conditions, this section studies the impact to adversarial success when
the examples are evaluated in the presence of three specific receiver effects,
which would likely occur during an OTA attack: AWGN, sample time offsets, and
center frequency offsets.

% ------------------------------------------------------------------------------
\subsection{Additive White Gaussian Noise} \label{sec:receiver:awgn}
AWGN has been shown to negatively impact both BER and classification accuracy. 
Additionally, as discussed in Section \ref{sec:performance:signalEffect:awgn},
AWGN can have a negative effect on adversarial success.  This section further
evaluates these negative effects with a larger scale study.  In some cases, such
as in ``rubbish examples'' \cite{RN68} or ``fooling images'' \cite{RN11}, the
primary goal of adversarial machine learning may simply be to create an input
that is classified with high confidence as some target class starting from a
noise input.  However, in general, fooling a classifier is a secondary goal that
must be balanced against the primary objective.  In CV, this primary objective
is to preserve human perception of the image.  In the current work, the primary
objective of self protect attacks is to transmit information to a friendly
receiver using a known modulation while the secondary objective is to avoid
recognition of that modulation scheme by an eavesdropper.  Therefore, this
section presents results showing the compounding impacts of adversarial machine
learning and AWGN on BER as well as the effect of AWGN on adversarial success
rates.

\begin{figure}
    \includegraphics[width=\linewidth]{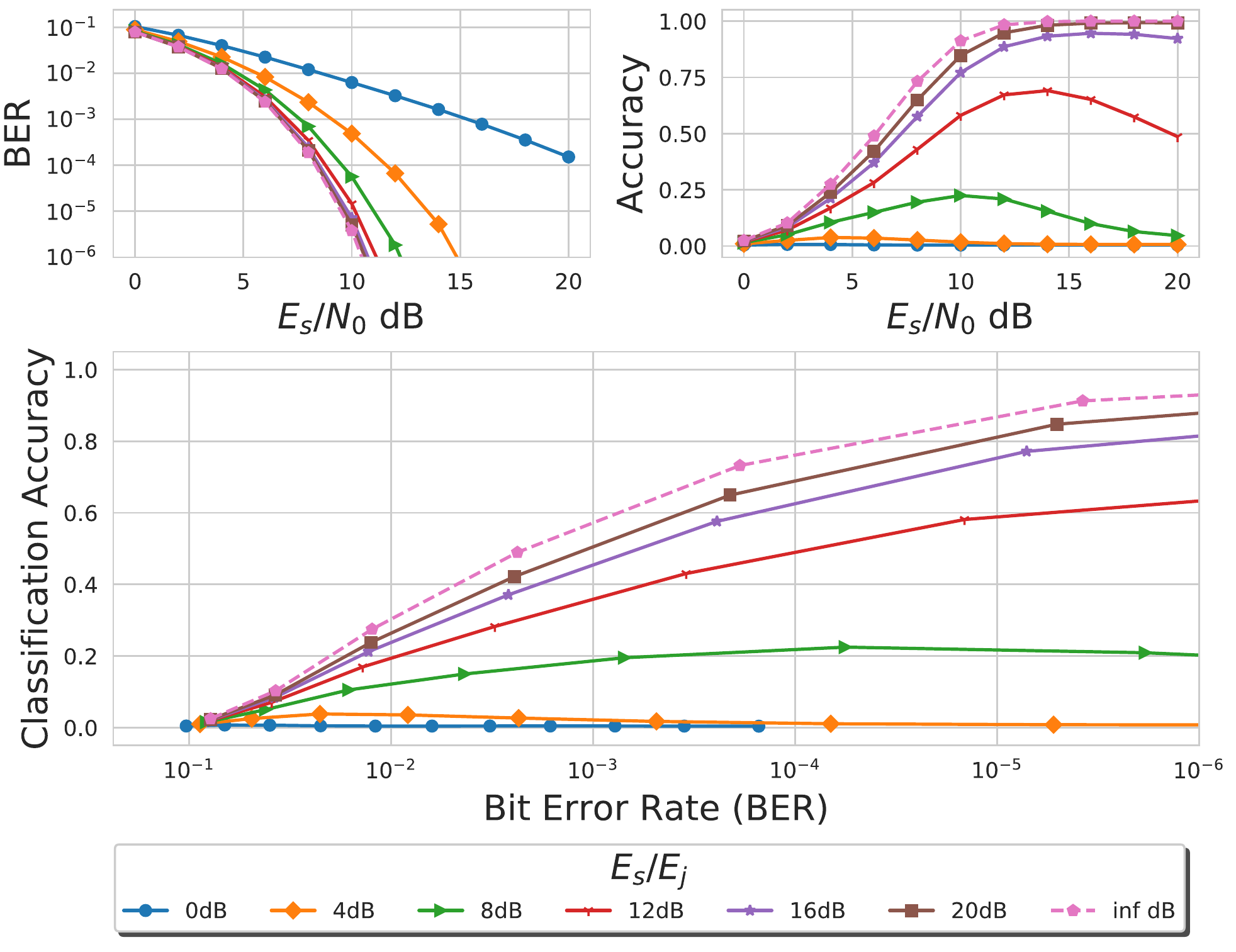}
    \caption{Classification accuracy and bit error rates at varying $E_s/E_j$ and $E_s/N_0$ for self protect untargeted adversarial attacks using FGSM on the model trained with Dataset A and a source modulation class of BPSK.}
    \label{fig:receiver:bpskAcc}
\end{figure}

\begin{figure}
    \includegraphics[width=\linewidth]{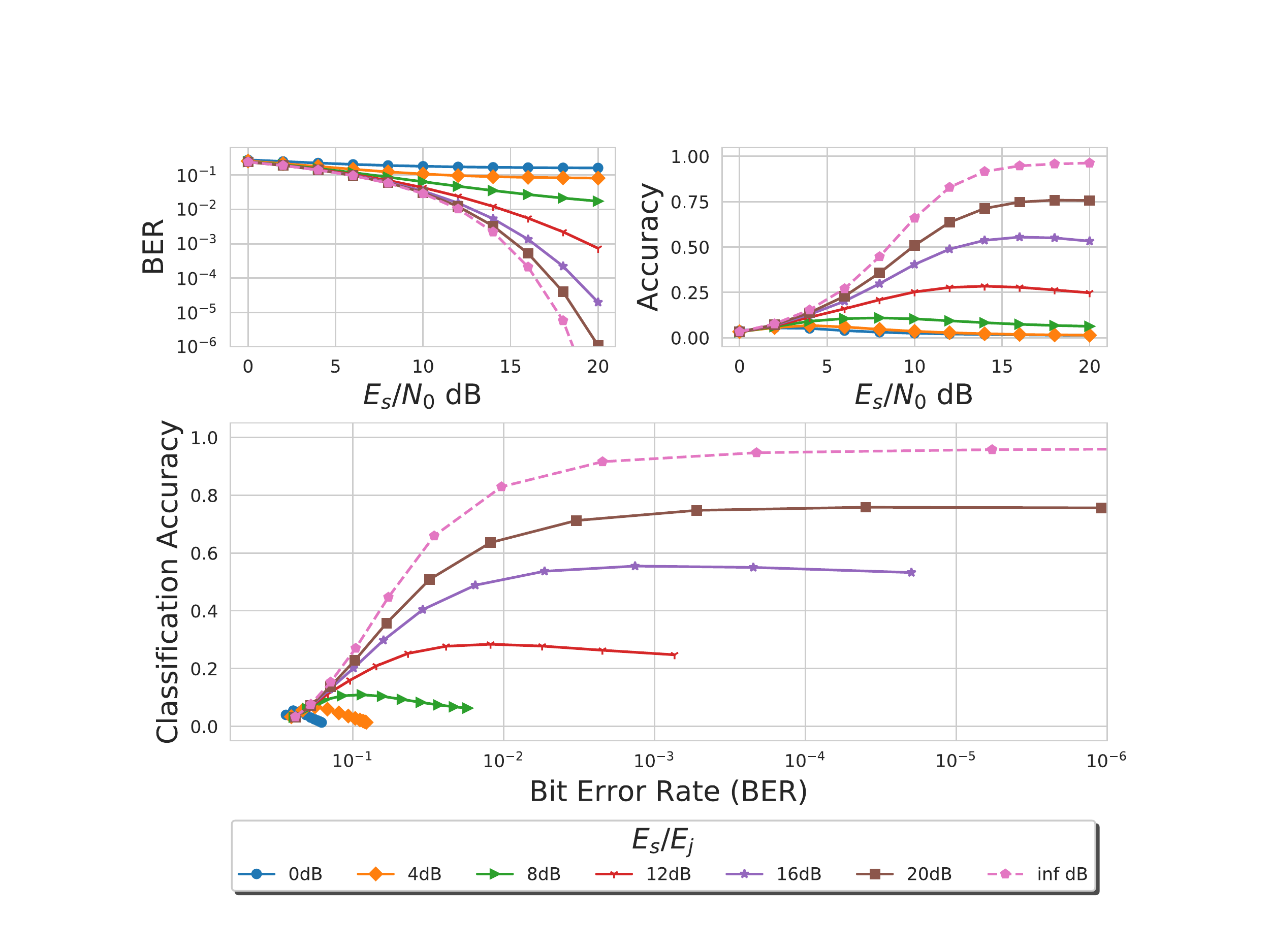}
    \caption{Classification accuracy and bit error rates at varying $E_s/E_j$ and $E_s/N_0$ for self protect untargeted adversarial attacks using FGSM on the model trained with Dataset A and a source modulation class of 8PSK.}
    \label{fig:receiver:8pskAcc}
\end{figure}

Using the model trained on Dataset A, a range of $E_s/N_0$ and $E_s/E_j$ are
considered.  For each $E_s/N_0$ considered, ten thousand trials are executed to
provide averaging of the random variables present in the channel model for a
given random signal.  The current work considers both the BER and classification
accuracy for BPSK in Figure \ref{fig:receiver:bpskAcc} and 8PSK in Figure
\ref{fig:receiver:8pskAcc}.

Unsurprisingly, increasing the adversarial perturbation energy has positive 
effects on adversarial success rates (also shown previously in Section 
\ref{sec:performance}) and negative effects on BER.  In order to directly
compare the trade space between the two across a range of SNRs, BER
versus classification accuracy is plotted for each $E_s/E_j$ considered.  At
high SNR, extremely low probabilities of bit error, such as those seen in BPSK
at $E_s/N_0=20$ dB, are hard to characterize empirically.  Therefore, in the BER
versus classification accuracy plots, all results with lower than $10^{-6}$ BER 
have been omitted for clarity.

By looking at Figure \ref{fig:receiver:bpskAcc}, one can observe that
classification accuracy can be degraded to $\approx 0\%$ with no noticeable
effect to BER for BPSK when using a white-box adversarial attack with an
$E_s/E_j$ of $4$ dB.  While this is a very strong result, it only occurs at high
SNRs ($>15$ dB). A more reasonable result to compare to would be the baseline
result at $10$ dB.  In order to achieve the same bit error rate as the baseline
of no attack (shown as a dashed line), an adversary must increase their SNR, and
therefore their transmission power, by $\approx 2$ dB when performing an
adversarial attack at an $E_s/E_j$ of $8$ dB.  A similar analysis can be
performed for QPSK where a $4$ dB increase to SNR is required to maintain the
same BER while reducing classification accuracy to $< 20\%$.

As stated in Section \ref{sec:performance:signalEffect}, AWGN can have negative
affects on adversarial success.  Therefore, while an eavesdropper with a high 
SNR would be fooled nearly all of the time by a BPSK transmission with an
$E_s/E_j$ of $8$ dB, an eavesdropper with an $E_s/N_0$ of $10$ dB would still
classify this signal correctly $20\%$ of the time.  If an adversary wished to
attain $0\%$ classification more generally for BPSK using FGSM, then they would
need to transmit with an $E_s/E_j$ of $4$ dB.  This attack intensity would
require an SNR increase of $\approx 4$ dB to maintain the same BER. 
The increased accuracy, at lower SNRs, observed previously in Figure
\ref{fig:performance:bpsk} can also be observed in Figure
\ref{fig:receiver:bpskAcc} and therefore generalizes across BPSK examples.  This
effect can also be observed, to a lesser extent, in the results of 8PSK (Figure
\ref{fig:receiver:8pskAcc}).  Additional experiments showed that the effect is
not observed for QPSK or QAM16.  Note that Figure
\ref{fig:performance:variance:qpsk} previously showed that the increased
sensitivity to noise for that QPSK example did not result in crossing the
decision boundary.  The effect of increased accuracy cannot be concluded from
QAM16 results because the baseline results already show a slight accuracy
improvement at SNRs around $10$ dB.

Evaluating attacker success in the case of higher order modulations such as 8PSK
and QAM16 is less clear.  Attacks with $E_s/E_j \leq 8$ dB already contain bit
errors without any added noise.  Therefore, degrading classification accuracy
of 8PSK below $20$\%, outside of the eavesdropper receiving the signal at low
SNR, would require forward error correction to account for the errors in
transmission.  In the case of QAM16, attacks using $E_s/E_j \leq 4$ dB would
impact the receiver more than the eavesdropper in many scenarios.  Specifically,
QAM16 has a BER of $\approx 16\%$ and $\approx 25\%$ when $E_s/E_j$ is $4$ and
$0$ dB respectively even when there is no additive noise.  Additionally, note
that both of these attack intensities are outside of the optimal range observed
in Figure \ref{fig:performance:qam16}.  Therefore, when evaluated as a function
of BER, the classification accuracy is actually lower in the baseline case than 
under the presence of these high intensity attacks.

In the case of QAM16, lower intensity attacks are effective at high SNR; 
however, they become less effective as SNR decreases, an anomalous effect 
previously discussed in Section \ref{sec:performance:signalEffect}.  Therefore,
untargeted adversarial machine learning with QAM16 as the source modulation
class may be most effective in situations where the eavesdropper is thought to
have a high fidelity capture of the transmission, such as when the eavesdropper
and transmitter are located in close proximity.  When the eavesdropper would
likely already have a weak view of the signal, it may be more effective to use
physical layer security concepts, such as lower transmission power or beam
steering, to further degrade the eavesdropper's signal capture.

These results conclude that adversarial machine learning is effective across
multiple modulations and SNRs to achieve the goal of untargeted  
misclassification because, for a given BER, classification can be
greatly reduced in many scenarios.  However, avoiding signal classification may 
require sacrificing spectral efficiency or increasing transmission power to
maintain the same bit error rate.  Additionally, AWGN was shown to have a
negative impact on adversarial success rates in 3 out of 4 source modulations 
tested and therefore adversarial machine learning can be the most effective at
high SNRs.

% ------------------------------------------------------------------------------
\subsection{Frequency Offset} \label{sec:receiver:frequency}
Signal classification systems typically do not know when and where a
transmission will occur.  Therefore, they must take in a wideband signal,
detect the frequency bins of the signals present, as well as the start and stop
times of transmission, and bring those signals down to baseband for further 
classification.  However, this process is not without error.  One effect shown 
in \cite{RN12} was the consequences of errors in center frequency estimation,
resulting in frequency offset signals.  The authors of \cite{RN12} found that
raw-IQ based AMC only generalized over the training distribution it was
provided and therefore if additional frequency offsets outside of the training 
distribution were encountered, the classification accuracy would suffer.
Because these estimations are never exact, adversarial examples transmitted over
the air must also generalize over these effects.

\begin{figure}
    \includegraphics[width=\linewidth]{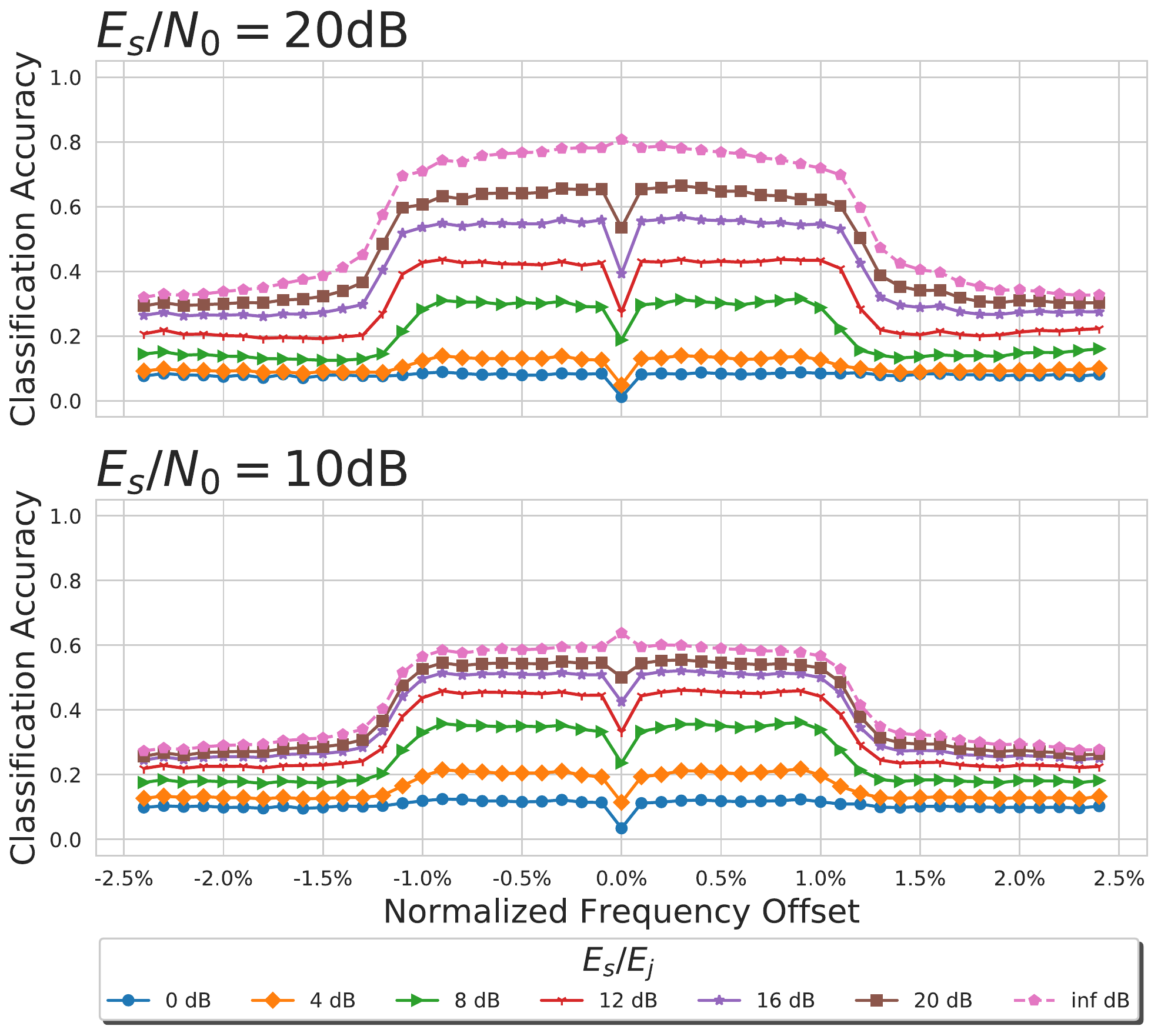}
    \caption{Classification accuracy vs normalized center frequency offset at varying $E_s/E_j$ for self protect untargeted adversarial attacks using FGSM. The model used is trained on Dataset B with an input size of $128$.  This dataset has training distribution of $\pm 1\%$ frequency offset that has been normalized to the sample rate.}
    \label{fig:receiver:freqOffsets}
\end{figure}

In order to evaluate the impact of center frequency offsets to adversarial
examples, it is necessary to use a model that has been trained to generalize 
over these effects.  Therefore, this experiment uses Dataset B, which has a
training distribution consisting of $\pm 1\%$ frequency offsets, which have been
normalized to the sample rate.  An input size of $128$ is used for closer
comparison to other results using Dataset A, which only has $128$ as an input 
size.  The frequency offsets are swept between $-2.5\%$ and $2.5\%$ with a step
size of $0.1\%$.  $E_s/N_0$ is evaluated at $10$ and $20$ dB.  At each SNR, 100
trials are performed to average out the effects of the stochastic process.  The
results of this experiment are shown in Figure \ref{fig:receiver:freqOffsets}.

It can be observed that the baseline classifier has learned to generalize over
the effects of frequency offsets within its training range of $\pm 1\%$;
however, the adversarial examples are classified with $\approx 10\%$ higher 
accuracy even at the lowest evaluated frequency offsets of $\pm 0.1\%$.  This
effect is observed at both $20$ and $10$ dB SNR.  Therefore, even minute errors
in frequency offset estimation can have negative effects on adversarial machine
learning and must be considered by adversarial generation methods.

% ------------------------------------------------------------------------------
\subsection{Time Offset} \label{sec:receiver:time}
\begin{figure}
    \includegraphics[width=\linewidth]{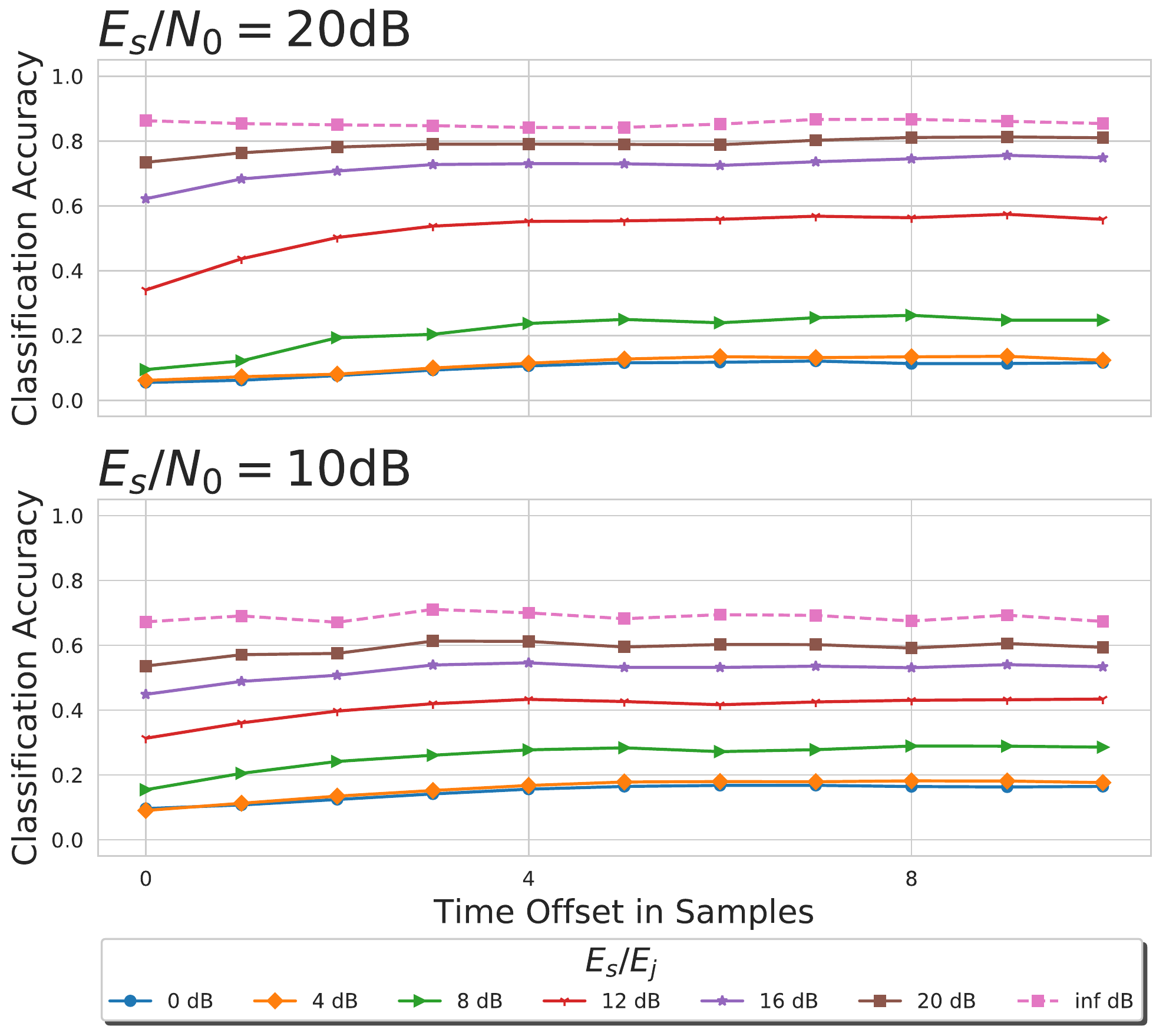}
    \caption{Classification accuracy vs time window offsets at varying $E_s/E_j$ for self protect untargeted adversarial attacks using FGSM. The model used is trained on Dataset A.}
    \label{fig:receiver:timeOffsets}
\end{figure}

An additional effect that could be encountered is sample time offsets.  In the
context of communications, sample time offsets can be thought of as a 
rectangular windowing function, used for creating discrete machine learning
examples, not aligning between the adversarial perturbation crafting and signal
classification.  As previously mentioned, the signal classification system must
estimate the start and stop times of a transmission; one way to estimate these
times is to use an energy detection algorithm where the power of a frequency
range is integrated over time and then thresholded to provide a binary 
indication of whether a signal is present.  A low threshold could have a high
false alarm rate and a high threshold could induce a lag in the estimation of
the start time.  Furthermore, signal classification systems could use
overlapping windows for subsequent classifications to increase accuracy through
the averaging of multiple classifications of different ``views'' of a signal or
use non-consecutive windows due to real-time computation constraints.  
Therefore, this effect is a near certainty.

This experiment uses the model trained on Dataset A and again evaluates the 
effect at an $E_s/N_0$ of $10$ and $20$ dB.  At each SNR, 100 trials are 
performed.  The time offset is modeled as a shift in the starting index used 
when slicing the signal for evaluating the signal classification performance and
non-overlapping/consecutive windows are still used.  The time offset was swept
from 0 to 127 (because the input size is 128 and this effect is periodic in the
input size); however, only the results from $0$ to $10$ are shown for 
simplicity.  Time offsets higher than $8$ samples, the symbol period, did not
present any significant additional impairments beyond those seen at $8$.  The 
results are shown in Figure \ref{fig:receiver:timeOffsets}.

As expected, the network is not heavily effected in the baseline case.  However,
the adversarial examples can be significantly impacted.  In the case of an
$E_s/E_j$ of 12dB, simply shifting the time window to the right by four samples
can increase the classification accuracy by $20\%$.  While some adversarial
perturbations have been shown to be agnostic to these time shifts, such as the
UAP \cite{RN80} attack considered in \cite{RN100}, all evaluations of 
adversarial machine learning in the context of RFML, that seek to model OTA
attacks, must assume this effect exists and generalize over it.

% ------------------------------------------------------------------------------
\section{Conclusions and Future Work} \label{sec:conclusion}
% -- All RFML Models Are Vulnerable --------------------------------------------
The current work has demonstrated the vulnerabilities of RFML systems to
adversarial examples by evaluating multiple example attacks against a raw-IQ
deep learning based modulation classifier.  First, it was shown that FGSM
\cite{RN68} crafted perturbations were vastly more effective than perturbations 
that were crafted using Gaussian noise at degrading the classifier accuracy when
the attack was launched with direct access to the classifier input.  Furthermore,
the current work demonstrated that these vulnerabilities were also present in
FGSM based OTA attacks by evaluating the attack effectiveness in the presence of
three RFML domain specific receiver effects: AWGN, sample time offsets, and
center frequency offsets.  When evaluating OTA attacks, evading an eavesdropper
is generally a secondary goal and must be balanced against the primary goal of
transmission, which is to communicate information across a wireless channel. 
Therefore, the current work showed that these attacks harmed the eavesdropper
more than the adversary by demonstrating that, for a given BER, classification
accuracy could be lowered for the majority of the OTA attacks considered.  Given
these results, it is logical to conclude that similar vulnerabilities exist in
all RFML systems when the adversary has white-box knowledge of the classifier.

% -- Future RFML Systems Must Adopt Defenses -----------------------------------
Future RFML systems must consider these vulnerabilities and develop defenses
against them.  The current work has shown that, while increasing the number of
samples used per classification can increase accuracy in the presence of AWGN,
it can also make the model more susceptible to adversarial examples.  Therefore,
future RFML systems could consider shrinking the input size at the cost of
accuracy in the baseline case.  Furthermore, the current work has reinforced the
viability of mutation testing \cite{RN38} by showing that RFML domain specific
receiver effects typically has a negative impact on adversarial examples. 
Consequently, using classifications from multiple views of the signal, with
different sample time offsets and center frequency offsets, can aid in detecting
adversarial examples and even properly classifying them.  However, RFML systems
are typically SWaP constrained and therefore increasing the number of inferences
per time step could limit the bandwidth that can be sensed in real time. 
Alternatively, defenses could be incorporated into the DNN training phase, which
is typically performed offline and thus has more computational resources or no
real-time processing constraint.  Ensemble adversarial training \cite{RN83} has
been shown as an effective method for hardening DNN models in the CV domain and
the results presented in the current work on BER penalties for adversarial
examples can be used to guide which examples to include during training.  RFML
does not necessarily need to classify all adversarial examples properly, but, it
could seek to balance an adversary's increasing success in evading the
eavesdropper versus their degrading ability to communicate information.

% -- Future Attacks Must Consider Noise/BER Impact -----------------------------
Future OTA adversarial evasion attacks must consider their ability to generalize
over RFML domain specific receiver effects as well as their their impact to the
underlying transmission.  The current work has demonstrated that all three 
receiver effects considered can degrade the adversary's ability to evade 
classification.  Furthermore, the current work has shown that, while current
adversarial methodology can be used for evading classification, especially when
using a lower order source modulation such as BPSK, it may require sacrificing
spectral efficiency or increasing transmission power to maintain the same bit
error rate.  Preliminary efforts in presenting additional adversarial methodology
may simply evaluate these effects, as we have done in this current work. 
However, more advanced efforts may directly incorporate these models of receiver
effects and wireless communications goals directly into their adversarial
methodology in order to create strong adversarial examples that generalize over
receiver effects and have limited impact to the underlying transmission.

% -- Final Sentence ------------------------------------------------------------
The current work concludes that adversarial machine learning is a credible and
evolving threat to RFML systems that must be considered in future research.

% ------------------------------------------------------------------------------
\bibliographystyle{ieeetr}
\bibliography{citations.bib}

% ------------------------------------------------------------------------------

\end{document}